# Conditional spinodal decomposition in Li-Mg anodes for lithium metal batteries

Leonardo Shoji Aota[a,b], Aubin Leray[c,d], Yuqi Liu[a], Frederic de Geuser[e], Chanwon Jung[a,f], Shyam Katnagallu[a], Tim M. Schwarz[a], Alisson Kwiatkowski da Silva[a,g], Júlio César Pereira dos Santos[g], Eric Marchezini Mazzer[h], Poonam Yadav[a,i], Christoph Freysoldt[a], Frank Stein[a], Yug Joshi[a], Se-Ho Kim[a,j], Dierk Raabe[a], Baptiste Gault[a,k]

[a] Max Planck Institute for Sustainable Materials (formerly Max-Planck Institut für Eisenforschung GmbH), 40237 Düsseldorf, Germany

[b] *present address:* Aeronautics Institute of Technology, 12228-900 São José dos Campos, Brazil

[c] Université de Picardie Jules Verne, 80039 Amiens, France

[d] Réseau sur le Stockage Electrochimique de l'Energie (RS2E), FR CNRS 3459, 80039 Amiens, France

[e] University Grenoble Alpes, CNRS, Grenoble INP, SIMAP, Grenoble F-38000, France

[f] *present address:* Department of Materials Science and Engineering, Pukyong National University, Busan 48513, Republic of Korea

[g] Thermo-Calc Software AB, Solna, Sweden

[h] Federal University of São Carlos, 13565-905 São Carlos, Brazil

[i] *present address:* Department of Metallurgical and Materials Engineering, Indian Institute of Technology Ropar, 14001 Punjab, India

[j] Materials Science and Engineering, Korea University, 02841 Seoul, Republic of Korea.

[k] Univ Rouen Normandie, CNRS, INSA Rouen Normandie, Groupe de Physique des Matériaux, UMR 6634, F-76000 Rouen, France

## Abstract

The development of batteries with high energy density, short charging times and use of sustainable materials is critical for decarbonization. Magnesium (Mg)-based anodes for lithium (Li) metal batteries promote homogeneous Li plating, thereby avoiding the formation of Li dendrites that cause short circuits and battery failure. However, microstructural modifications induced by Li-alloying and their influence on battery operation remain elusive. Here, we unveil the previously unknown formation of an ordered B2 phase, which creates a conditional spinodal decomposition with the β-body-centered cubic phase. Chemical fluctuations characteristic of spinodal decomposition give rise to uniformly dispersed Li-rich β-BCC and Li-poor B2 continuous interconnected phases, with the former providing a fast diffusion pathway for Li diffusion towards the anode, hence decreasing the propensity for dendrite formation at elevated current density. This is achieved using Earth-abundant and inexpensive Mg.

The development of batteries with both higher energy densities and shorter charging times, while employing abundant and cost-effective materials to facilitate sustainable scalability[1], is required to decarbonize modern society. Nowadays, graphite anodes dominate commercial applications; yet they offer a limited gravimetric capacity[2] of 372 mAh/g, which is translated into limited energy density, and an increased risk of supply disruption[3]. Metallic Li is theoretically the ideal anode due to its gravimetric capacity of 3860 mAh/g and the lowest electrochemical potential (-3.04 V vs. standard hydrogen electrode[4]) among all anodes. However, uncontrolled Li growth in protruding topological form, referred to as dendrites[5], along with the high reactivity of metallic Li with the electrolyte[6], cause poor cycling stability and thermal runaway. Alloying anodes can circumvent these issues while still providing attractive gravimetric capacity, and hence, high energy density in a battery.

Li-Mg alloys appear as versatile candidates to be used as anodes for lithium batteries. In currently accepted thermodynamic databases for this system[7], only two phases are expected at room temperature: α-hexagonal close packed (α-HCP) (from 0 to 17 at.% Li) and β-body centered cubic (β-BCC) from 33 to 100 at.% Li. Only the BCC phase is beneficial for anodes due to the absence of the following three factors in the α-HCP phase.

First, the β-BCC phase has a high effective bulk Li diffusivity of $10^{-7}$ – $10^{-10}$ $cm^2/s$[8–10] and a high surface Li diffusivity[11–13]. Together, these create a homogeneous Li flux and current density distribution[14], with non-localized nucleation sites[15], avoiding formation of dendrites. Second, the interface stability between the anode and the electrolyte is improved when using Li-Mg alloys compared to pure Li[16–19] due to decreased reactivity toward the electrolyte[18]. Third, the large Li solid solubility in the β-BCC phase down to pure BCC-Li reduces the nucleation overpotential for Li plating due to the similarity of crystal structure and lattice parameters between the anode and the plated Li[11,20–22].

Here, we demonstrate that the β-BCC solid-solution phase-field range is not as large as conventionally determined to be ca. 33 – 100 at.% Li. Cryogenically-enabled atom probe tomography (APT)[23,24] reveals compositional fluctuations in this phase-field region with a Li-rich (> 84 at. %Li) β-BCC_A2 disordered phase and its Li-poor (<62 at.% Li) BCC_B2 ordered counterpart at room temperature. Transmission electron microscopy (TEM) and differential scanning calorimetry (DSC) confirm long-range ordering reaction. A miscibility gap separates the Li-rich β-BCC and Li-poor B2 phases, with ordering being necessary to trigger a so-called conditional spinodal reaction[25,26]. This means that the formation of the ordered B2 phase is a prerequisite for spinodal decomposition. Without B2 ordering, the decomposition would not occur.

In the present case this conditional spinodal decomposition occurs homogeneously in the bulk, enabling a continuous nanoscale 3D path for fast diffusion of Li, as the Li diffusivity scales with the Li content. Such drive for the consumption of the plated Li away from the anode's surface decreases the propensity for dendrite growth. These insights reveal the previously unknown complex phase transformation behavior in the Li-Mg system, which directly affects the electrochemical behavior and alter the mechanical properties during battery charging/discharging. These phase transformations could be used to engineer the microstructure of the anode towards optimized chemomechanical, and hence improved cycle stability in a battery.

## Results

### Long-range ordering in the pristine alloy

A 28.9 Li-71.1 at.% Mg alloy was prepared with the composition confirmed by atomic absorption spectroscopy (AAS) after casting and solidification of the alloy in a Cu mold (see Methods). The sample was stored at room temperature for over 14 years. By combining, electron backscatter diffraction, X-ray diffraction and backscattered scanning-electron micrographs in Figures S1 and 1a, respectively, we hypothesize a darker β-BCC matrix, along with dispersed α-HCP grains, see Supplementary Discussion. A strong preferential crystallographic orientation is obtained from solidification.

Specimens from the α-HCP grains (lighter grains in Figure 1a) were prepared for APT and TEM by using a focused-ion beam technique (FIB)[27]. In the APT reconstruction displayed in Figure 1b, a set of 25 at.% Li isoconcentration surfaces highlights precipitates within the α-HCP grain. The α-HCP grain has a concentration of 16 at.% Li that lies close to the Li solubility limit, whereas nearly spherical precipitates with a diameter around 10 nm contain around 50 at.% Li, Figure 1c, deviating from the expected concentration in current phase diagrams.

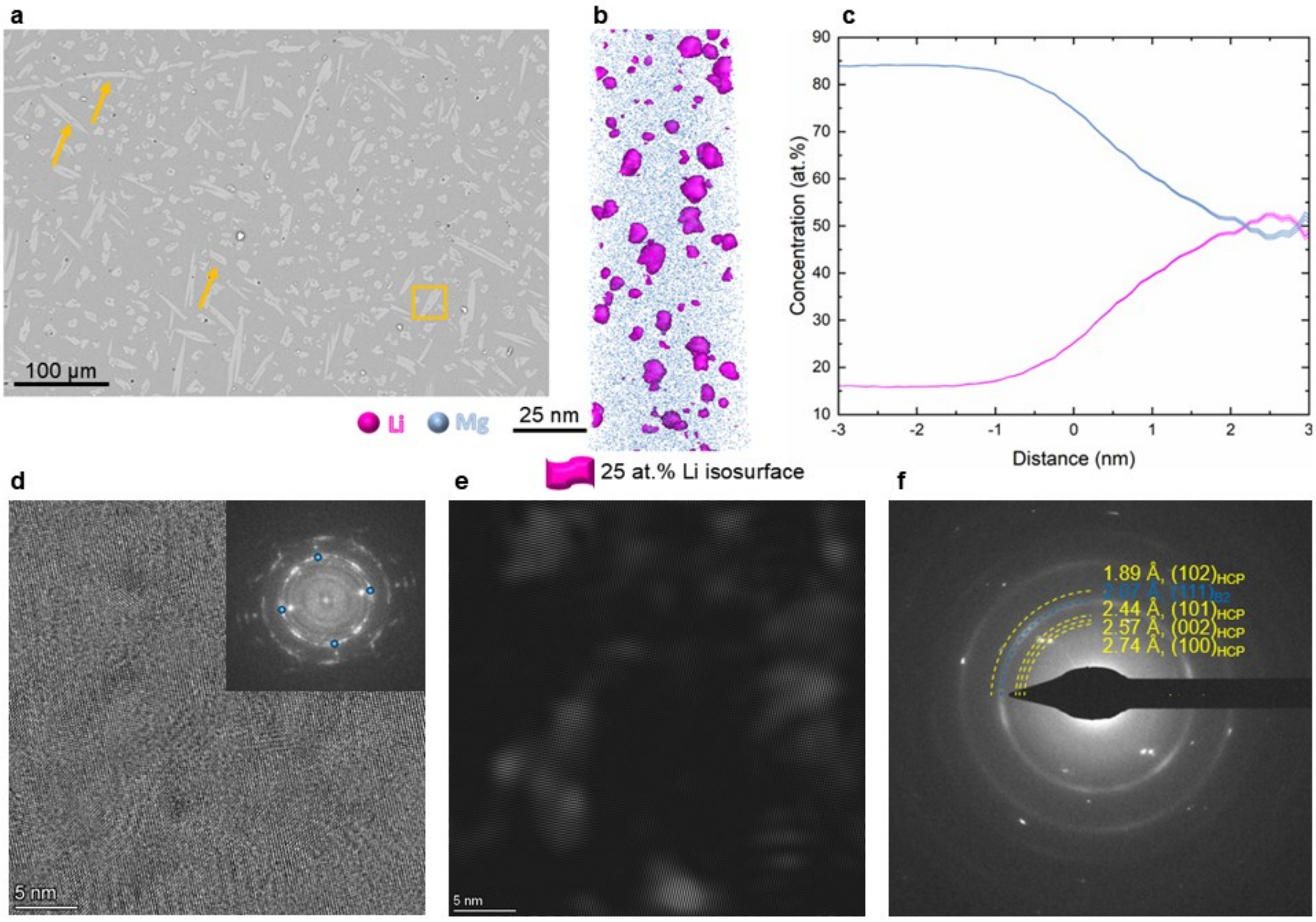

**Figure 1 –** (a) Backscattered scanning electron micrographs of the pristine Li-71.1 at.% Mg alloy. The dark gray matrix is the β-BCC phase. The light gray regions are α-HCP are richer in Mg (poorer in Li). Orange arrows indicate finer α-HCP grains both within the matrix and along grain boundaries. (b) Section from an atom probe tomography reconstruction from the pristine alloy with 25 at.% Li isoconcentration surfaces highlighting the Li-rich precipitates within the α-HCP grains, as indicated by the orange box in (a). (c) Composition profile as a function of the distance to the set of 25 at.% Li isoconcentration surfaces, showcasing a Li content near 50 at.% in the core of the Li-rich precipitates, which deviates from the composition expected for the pristine alloy based on the equilibrium phase diagram. (d) high resolution-transmission electron microscopy image of the pristine Li-71.1 at.% Mg alloy with nanoscale precipitates, the corresponding fast Fourier transformation as inset. The B2 diffraction spots, selected for the image (e) are shown as blue circles. (e) Selecting only the B2 diffraction spots from the fast Fourier transformation taken from (d), reveals their spatial distribution in the HR-TEM image. (f) Diffraction spots from both HCP and B2 crystal structures obtained from a region of interest using 400 nm-diameter selected area electron diffraction aperture.

High-resolution transmission electron microscopy, Figure 1d, evidence such precipitates as BCC_B2 phase, highlighted by selecting a spot in the fast-Fourier transform of the image corresponding to an interplanar distance of 2.07 Å in Figure 1e (see Methods). This structure is confirmed by selected-area electron diffraction, Figure 1f. Although the formation of an ordered B2 crystal structure in the Li-Mg system had not been experimentally observed until now[28], it had been previously suggested by first-principles calculations[29,30] and appears thermodynamically plausible, based on the negative enthalpy of mixing of this phase in Figures S2a-b (see Supplementary Discussion). To confirm the ordering reaction under thermodynamic equilibrium, a separate 50-50 at.% of Li alloy was cast, homogenized (see Methods), and analyzed by DSC, detecting a reversible B2 → BCC first-order phase transition at 390 ± 3°C in Figure S2c.

The lack of previous experimental observation of this B2 phase in the Li-Mg system may stem from the limited spatial resolution and detection difficulties of light elements, such as Li[31], faced by conventional characterization techniques. In addition, degradation upon sample preparation, air transfer and beam damage during characterization can take place[32,33], challenges avoided in our study by using cryogenically enabled specimen preparation approaches

**Conditional spinodal decomposition and its implications on the phase transformations during lithiation**

To sweep the phase diagram towards the Li-rich side, the Li28.9 alloy was lithiated at 20 μA/cm$^2$ until 6 mAh/cm$^2$, followed by storing in an Ar glovebox for 3 weeks at room temperature to homogenize the composition gradient along the depth and proceed towards equilibrium conditions.

Cryo-APT analyses (see methods) were performed at different depths 10 – 15 μm below the surface, from the microscopically uniform dark gray region readily visible in the backscatter electron scanning electron micrograph in Figure 2a. A set of isoconcentration surfaces in an APT reconstruction, Figure 2b, evidences a pattern of alternating Li-rich and Li-poor regions characteristically attributed to spinodal decomposition[34,35], further showcased in the compositional map in Figure 2c. The average Li concentration is 62 at.% Li (see also Figure S3). A composition profile calculated as a function of the distance to the set of 60 at.% Li isoconcentration surfaces (proximity histogram[36]) is plotted in Figures 2d, with Li-poor and Li-rich regions containing 56 and 77 at.% Li, respectively. The possibility of phase transformations below room temperature[29], i.e. during cooling to 50 K (-223°C) in the atom probe, was discarded based on cryogenic DSC (see Figure S4 and Supplementary Discussion). We can infer that Li-poor regions are B2-ordered, whereas Li-rich regions remain β-BCC, supporting the appearance of a conditional spinodal decomposition[25,37,38]. This allows us to rationalize the contrast observed in Figure 2a, with the darker contrast corresponding to the β+B2 region and the lighter to α+B2 region, indicating that beyond approximately 60 μm depth the Li composition is below the β-BCC stability range.

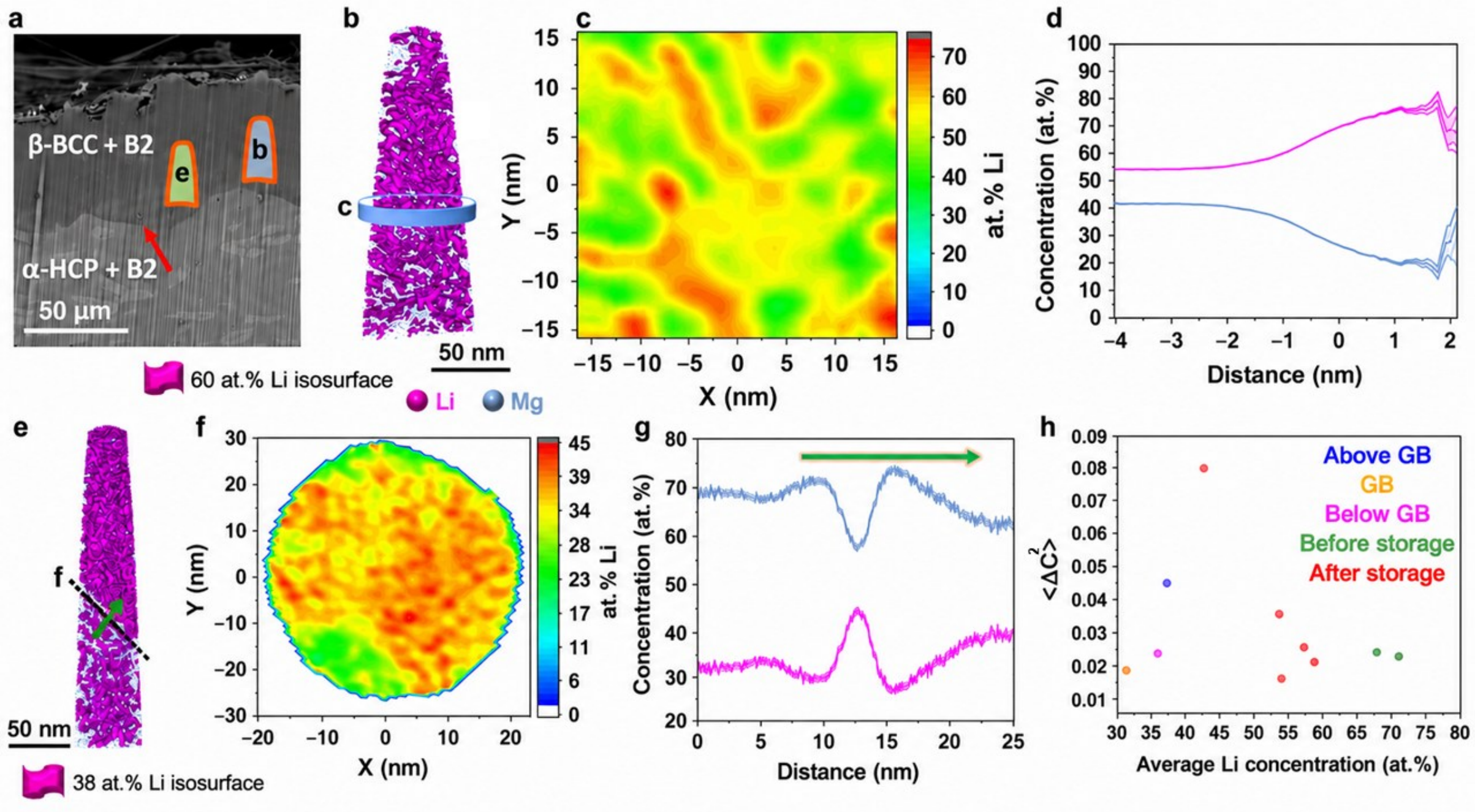

**Figure 2 –** Atom probe tomography (APT) reconstructions from a galvanostatically lithiated 28.9 at.% Li sample under 20 μA/cm$^2$ until 6 mAh/cm$^2$ areal capacity. The sample was stored for 3 weeks at room temperature under Ar. (a) The cross section imaged at room temperature with backscattered electrons (BSE) indicates from where the atom probe samples were taken at different depths inside the Li-rich topmost layer. The top surface corresponds to the anode surface in contact with the electrolyte. (b) APT reconstruction from the top region, at approximately 10-15 μm depth, exhibiting spinodal fluctuations highlighted by 60 at.% Li isoconcentration surfaces. (c) Li concentration heat maps taken from the 5 nm thick disk in (a), further evidencing the formation of compositional fluctuations. (d) Proximity histogram calculated from the 60 at.% Li isoconcentration isosurface from (b), with Li-rich and Li-poor regions having 56 and 77 at.% Li. (e) APT reconstruction at approximately 50 μm depth, still in the dark gray area, indicating a lower Li content, with spinodal fluctuations in both grains and at the grain boundary. (f) Compositional fluctuations also appearing in Li concentration heat maps at the grain boundary indicated in (e). (g) The grain boundary is enriched in Li and the regions adjacent to the boundary are depleted in Li in the 1D chemical concentration profile. (h) Chemical composition amplitude $<\Delta C^2>$, the difference between the Li concentration in Li-rich and Li-poor regions in fraction, as function of the average Li concentration. In (h), above GB, GB and below GB stand for the corresponding regions of interest from the dataset shown in (e).

At higher Li concentration, near the top surface, a homogeneous matrix and Li-rich precipitates larger than 20 nm and reaching up to 90 at.% Li are imaged by APT in Figure S5. Additional dark features indicated by red arrows in the electron micrograph in Figure S6 correspond to micrometric β-BCC-Li precipitates with a Li composition above 99.5 at.%. Although APT analysis of pure Li is known to be challenging[39], impurities within the Li may help stabilize the field evaporation process[40], yielding reliable quantification (see SI).

Unlike classical nucleation, which requires a large local composition fluctuation to overcome an energy barrier, spinodal decomposition initiates from very small-amplitude composition fluctuations spread over a large volume[41,42]. Consequently, Li-rich and Li-poor regions with the same crystal structure form spontaneously, without an energy barrier[42] (see Supplementary Discussion). A conditional spinodal occurs when the formation of an ordered phase produces a miscibility gap. This ordered structure (BCC_B2) becomes unstable under Li enrichment, due to a limitation for the compositional deviation from stoichiometry, favoring the formation of its disordered counterpart phase (Li-rich β-BCC) in an alternating pattern shown in Figure 2c.

Thus, the conditional spinodal decreases the nucleation barrier of Li-rich β-BCC by. The B2-ordering and decomposition occur concomitantly during ageing, with the evolution of one phase transformation accelerating the other[37]. Further lithiation causes a transition from the spinodal fluctuations into a "two-phase" (Li-rich β-BCC and Li-poor B2) microstructure, similar to that obtained by nucleation and growth, with two chemically homogeneous regions and a sharp interface[41,42], as observed in Figures S5-6 and discussed in Supplementary Information.

**Effect of microstructure on the local phase transformations during lithiation**

The grain boundary from Figures 2e evidences Li enrichment along with the presence of spinodal fluctuations along the grain boundary plane, i.e. periodic alternating Li-rich and Li-poor region in Figure 2f. These two observations suggest that grain boundary Li segregation precedes the spinodal state by thermodynamically driven grain boundary segregation[43,44]. The Li-poor regions on either side of the grain boundary (25-28 at.% vs. 43 at.% Li at the grain boundary) from Figure 2g, are akin to precipitation-free zones (PFZ)[45]. Li segregated to grain boundaries leave the abutting region depleted in Li, because the faster Li transport down the network of grain boundaries prevents the replenishing of Li in these regions. As a result, the Li composition in regions surrounding grain boundaries remain outside or near the limit of metastability, thereby precluding the initiation of the spinodal decomposition.

Due to the Li enrichment, grain boundaries are already in the second stage of spinodal decomposition after 3 weeks of storage, while the adjacent grains are in the first stage, as suggested by the higher wavelength for the former (orange icon in Figure S3b, discussed in Supplementary Information). The amplitude of the chemical fluctuations, $<\Delta C^2>$ were calculated based on radial distribution functions[43,44] (see Methods and Figure S3 for more information), and it does not vary systematically between grain boundaries and other regions, even after 21 days storage, in Figure 2h. The local Li enrichment at grain boundaries will drive the earlier nucleation of Li-rich β-BCC at grain boundaries and their faster growth further enhancing its local Li diffusivity, as the Li concentration in the Li-rich region scales with the average Li concentration given the same chemical amplitude in Figure S3d.

Overall, the observed onset of spinodal decomposition and precipitation of Li-rich β-BCC accelerated at grain boundaries suggests a strong role of the alloy's initial grain size and the distribution of grain boundary misorientation. Engineering the microstructure, such as reducing the grain size, may offer a suitable path to optimize the anode performance in terms of capacity retention and rate capacity.

**Kinetic effect on conditional spinodal decomposition**

The net driving force for Li insertion is controlled by the electro-chemical potential. The rate or flux of Li ingress into the electrode is controlled by diffusion mediated by chemical potential gradient. For compositions near the limit of metastability, spinodal decomposition may be kinetically hindered and superseded by a nucleation and growth mechanism[46] in case the Li diffusion within the alloy is fast enough[47].

To study which mechanism occurs at higher current densities, the 28.9 Li alloy was lithiated at 25 μA/cm$^2$ and 200 μA/cm$^2$. The galvanostatic lithiation results in a drop of potential close to 0 V (vs. $Li^+$/Li) after 0.8 mAh/cm$^2$ under a current density of 25 μA/cm$^2$, while the potential goes directly to negative values vs. $Li^+$/Li at 200 μA/cm$^2$ in Figure 3a. At 200 μA/cm2, the lithium nucleation and growth potentials[48,49] are -90 and -23 mV vs. ($Li^+$/Li), respectively, whereas at 25 μA/cm$^2$ only a distinguishable Li growth potential is observed at -7 mV vs $Li^+$/Li. The plated metallic Li layer at 25 μA/cm$^2$ cannot be distinguished from the underlying anode in Figure 3b due to sufficient time for Li diffusion inside the Li-Mg alloy at lower current densities. Correspondingly, samples taken from the very top do not correspond to 100 at.% Li average concentration, Figure 3c.

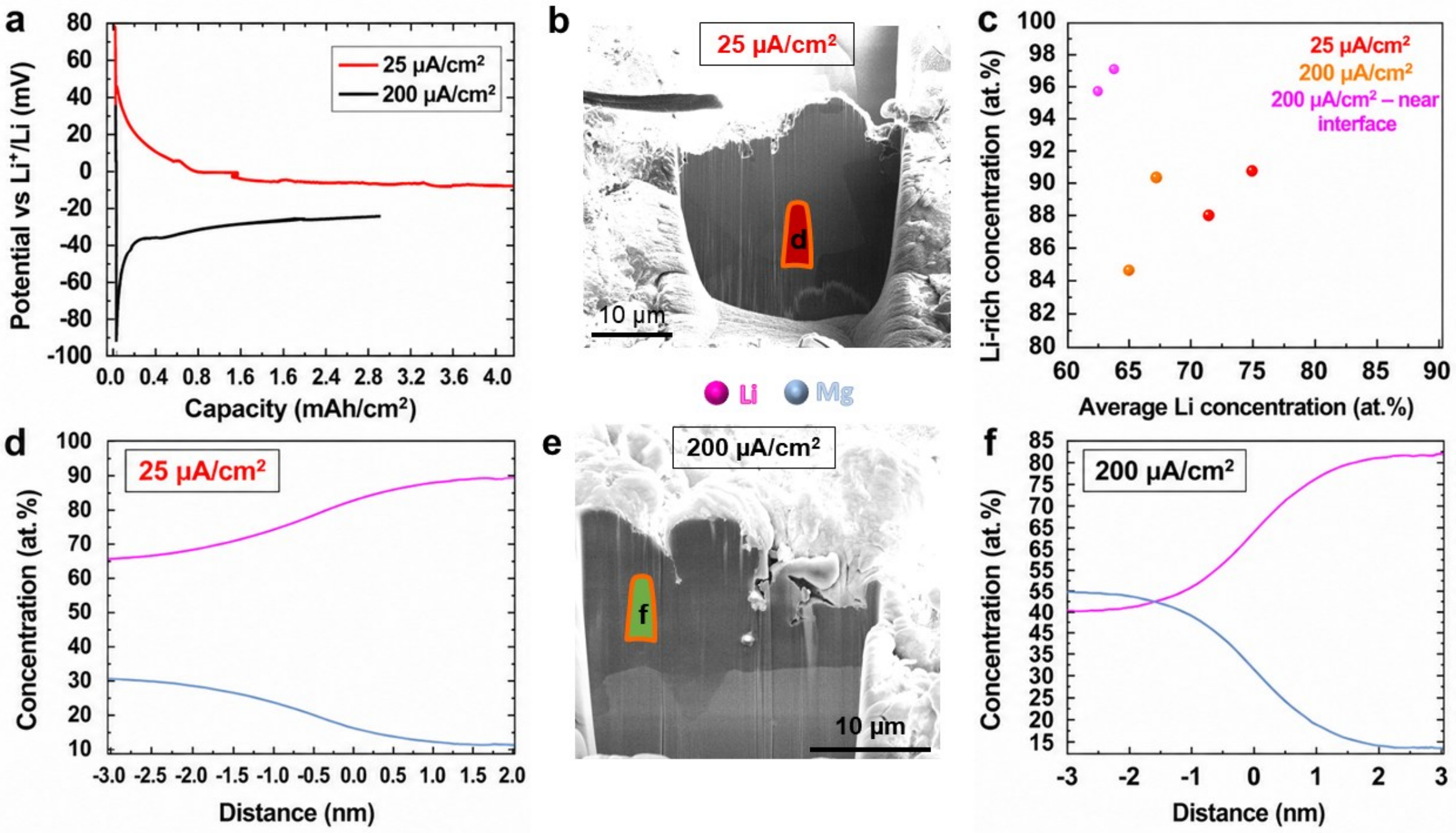

**Figure 3 –** (a) Voltage-capacity curves for galvanostatic lithiation under two different current densities for Li-71.1 at.% Mg alloy at 25°C. (b) Cross section of the sample lithiated at 25 μA/cm$^2$, showing an equiaxial grain structure under the ion beam imaging. (c) Chemical amplitude <ΔC$^2$> taken calculated based on the DIAM and PCF methods for samples immediately analyzed after lithiation under 25 and 200 μA/cm$^2$. There is not a clear trend, except a higher chemical amplitude at the interface between dark and light gray regions shown in (e). (d) Proximity histogram obtained from the sample lithiated at 25 μA/cm$^2$, and using a 79 at.% Li isoconcentration surface, with Li-rich regions reaching around 88 at.% Li. (e) Cross section of the lithiated sample under 200 μA/cm$^2$, where two layers are visible. The location of the atom probe tomography reconstruction from where the proximity histogram was taken is schematically shown. (f) The proximity histogram obtained with a 60 at.% Li isoconcentration surface, with Li-rich and Li-poor reaching 83 and 47 at.% Li, respectively.

In the anode lithiated under 25 μA/cm$^2$ imaged in Figure 3b is marked a region selected for APT analysis at a depth of 10 – 15 μm. The corresponding proximity histogram taken from 79 at.% Li concentration isosurfaces is plotted in Figure 3d, with Li-poor B2 regions containing 65 and Li-rich BCC regions with 88 at.% Li. By increasing the lithiation current density to 200 μA/cm$^2$, around 5 μm below the top surface in Figure 3e, the spinodal decomposition is once again observed, with an average Li concentration of 66.9 at.% as shown in Figure 3f. Li-rich regions reach 83 at.% Li, whereas Li-poor regions have around 47 at.% Li. Despite the difference in current density, there is not a clear trend regarding the chemical amplitude, correlation length or wavelength, which does not allow us to distinguish between the stage of spinodal decomposition in Figure S7. The bottom region below the interface in Figure 3e has an average Li content of approximately 33 at.% Li, close to the pristine alloy in Figure S8.

**Discussion**

The spinodal decomposition hence occurs over the whole studied range of current densities, given that Li diffusivity in B2 phase is insufficient to overcome the spinodal decomposition, i.e., sweep the whole unstable composition range prior to the onset of spinodal. Yet, the chemical composition profile still changes depending on the applied current density. Based on current thermodynamic databases, the phase diagram and the expected Li profile are represented in Figurea. The Li compositional gradient would be lower for the β-BCC phase at the top due to a higher Li diffusivity when compared to the α-HCP + β-BCC underneath. With increased Li ingress, the volume fraction of β-BCC phase would increase in such microstructure under constant Li concentration in both phases.

The presence of conditional spinodal decomposition differs from the α+β microstructure. We modelled the Li-Mg system using the order–disorder formalism[50], to describe a BCC_B2 phase using a four-sublattice thermodynamic model (see SI). Despite the A2→B2 transition being historically reported as a second-order phase transformation in most systems, the Mg-Li system undergoes such transition as first order because of the employed four-sublattice thermodynamic model. Although a more detailed assessment is required for future work, the resulting phase diagram (Figure 4b) reflects the experimental observations reported here, including the first order transition in Figure S2c, and is thermodynamically consistent, see Figure S9.

In the new phase diagram, a two-phase α+B2 region is predicted in the pristine alloy, and lithiation triggers the evolution of the microstructure towards B2, followed by another two-phase B2 + β-BCC region via spinodal decomposition, and finally, a single β-BCC phase at higher Li content. At Li concentrations above 90 at.% Li, only β-BCC exists in the microstructure, such as near the electrode surface.

Even at 25 μA/cm$^2$, the potential at the surface reaches negative values after 0.8 mAh/cm$^2$ in Figure 3a, indicating the plating of metallic Li, i.e. the Li diffusivity in β-BCC is not high enough to move sufficient Li from the surface towards the anode interior and decrease the Li activity at the surface. Instead, the Li activity reaches 1, and thus Li plating acts as alternative mechanism to reduce Li on the anode side. We suggest that at 25 μA/cm$^2$ mostly alloying takes place, and Li is plated as a discontinuous layer at the surface in Figure 4c. By increasing the current density (≥ 200 μA/cm$^2$), more Li is plated at the top of the anode, dominating the lithiation behavior. A continuous metallic Li layer is produced, which allows a superior Li supply into the anode while keeping the surface Li content at 100 at.% in Figure 4d. The result is a higher Li content near the surface of the anode than at lower current densities.

Further increasing the current density, the anode chemical composition profile will be the same regardless of the employed current density, only with a thicker Li layer formed at the surface, due to the kinetic limitation from Li diffusivity in the anode. However, dendrite growth will take place above a critical current density[5]. As soon as a continuous plated Li layer is formed on top of the anode, the Li-anode system behaves identically as a diffusion couple, where Li alloying is driven purely by chemical potential. To showcase this example, micrometric pieces of pure Mg and pure Li were contacted within the FIB-SEM and reacted for 14 h at room temperature in the absence of any external potential (Figure S10a). The same cryogenic-UHV transfer and atom probe analysis routine was performed in this micro-diffusion couple[51]. The conditional spinodal decomposition is again observed with similar Li content in the proximity histogram, i.e., B2 with 43 at.% Li vs. β-BCC with 76 at.% Li in Li-poor and Li-rich regions, respectively, in Figures S10b-c.

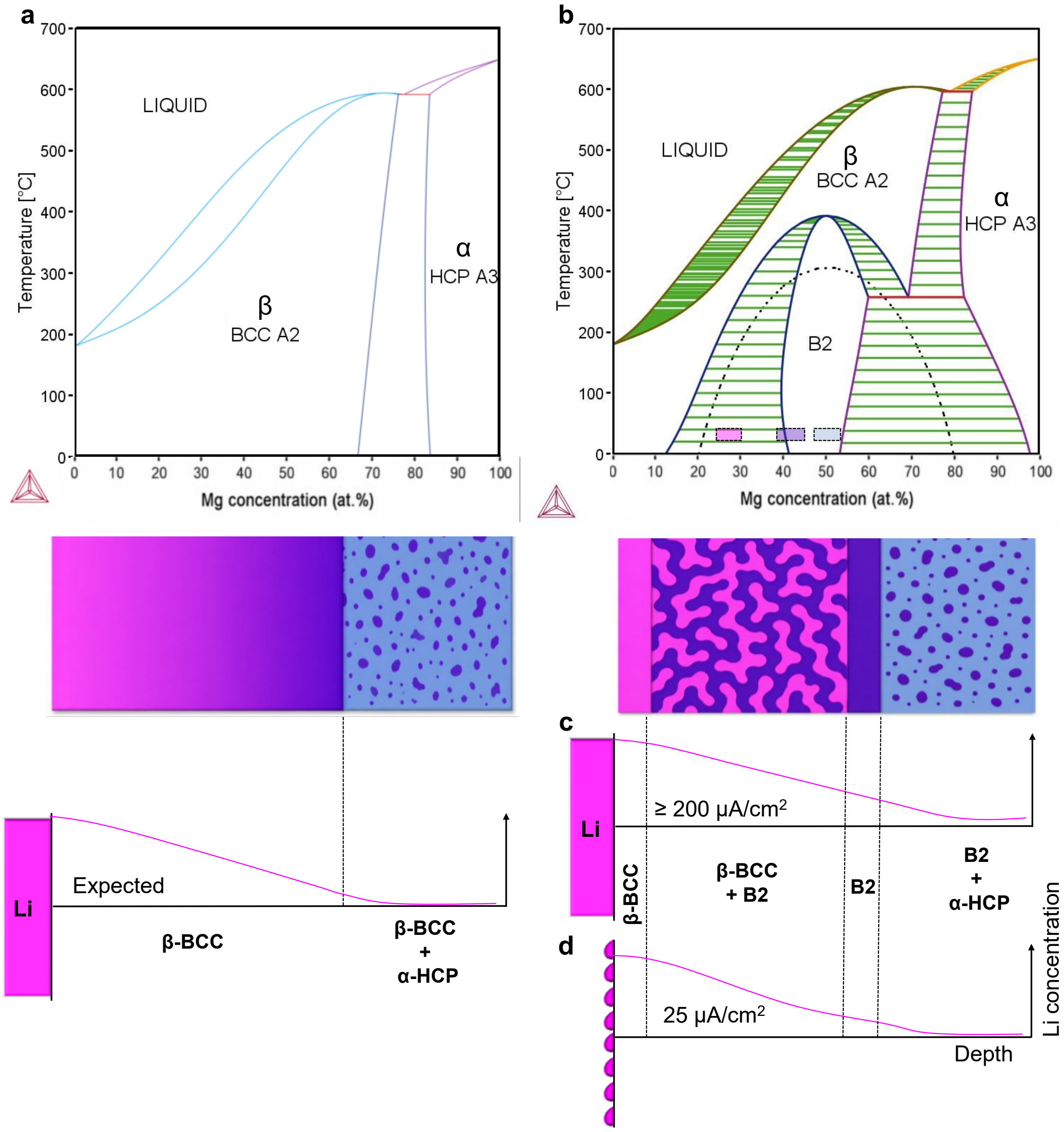

**Figure 4 –** Li concentration profile along the depth as function of the current density. (a) Calculated Li-Mg phase diagram based on current databases, with the corresponding expected concentration profile from equilibrium thermodynamics, where a single phase is observed at Li > 33 at.% Li and a two-phase microstructure (α-HCP + β-BCC) is observed in the Li-poor region. The Li chemical composition gradually reaches the composition of the pristine sample at deeper regions. (b) The Li-Mg phase diagram calculated with the thermodynamic parameters assessed in this work (see Methods) with an ordered B2 phase. The purple, light blue and magenta rectangles in the phase diagram represent the chemical composition range from the Li-rich regions, average chemical composition and Li-poor regions, respectively, taken from atom probe tomography measurements after lithiation at 25 μA/cm$^2$ and 3 days storage, as taken from Figure S3. Green regions correspond to two-phase regions in the phase diagram. (c) The magenta Li semicircles at the top represent discontinuous plated Li nuclei at the surface of the anode at 25 μA/cm$^2$, whereas (d) a continuous metallic Li layer is plated for current densities ≥ 200 μA/cm$^2$. Such continuous layer may be effectively present only as a nanometric layer at the top surface, creating a continuous surface composition of 100 at.% Li. The amount of Li near the surface,

inside the anode, and a larger chemical gradient, is observed at higher current densities, where the surface composition of the anode is kept constant at 100 at.% Li. The magenta to blue color scale in the illustration indicates the local Li or Mg-enrichment, respectively, characterizing the observed microstructures.

The conditional spinodal decomposition affects the electrochemical behavior of the anode. It thermodynamically assists and may even accelerate the formation of Li-rich β-BCC within the anode itself[37], starting from the top and advancing throughout the depth while consuming the plated Li. The Li diffusion occurs mainly via a continuous nanoscale 3D path for fast diffusion of Li within Li-rich β-BCC region created by spinodal fluctuations, as the Li diffusivity scales with Li concentration[52]. The synergistic acceleration of both ordering and spinodal fluctuations occurs throughout the bulk, further reducing the propensity for dendrite formation.

For Li-Mg alloys with Li content above 90 at.%, conditional spinodal decomposition may not take place during the first lithiation. However, a phase separation will take place during the following delithiation at commercially relevant current densities due to the local Li depletion near the surface[40,53]. The local Li depletion is caused by insufficient Li diffusivity to follow the thermodynamic equilibrium. Therefore, even when using pristine alloys with > 90 at.% Li, the Li concentration can locally reach the miscibility gap or the B2 phase field, possibly offering a kinetic barrier for Li diffusion, leading to Li trapping, once the B2-LiMg phase is formed. This spinodal decomposition is hence relevant for all ranges of composition for Li-Mg alloys and all current densities.

In summary, a conditional spinodal decomposition was revealed between B2 and β-BCC phases from the Li-Mg system, allowing the nucleation of Li-rich β-BCC from Li-rich fluctuations with negligible energy barrier within the bulk anode. The conditional spinodal decomposition likely takes place under all current densities given the insufficient Li diffusivity in the B2 and β-BCC phases to shift the local chemical composition outside the metastability limit prior to the spinodal decomposition onset. Grain boundaries offer sites for accelerated precipitation in such alloys, due to a combination of grain boundary segregation and diffusivity. The formation of Li-rich β-BCC regions by spinodal fluctuations within the bulk anode is possibly an additional mechanism to reduce the propensity for dendrite nucleation and growth in Li-Mg alloys. However, the formation of B2 phase during delithiation may impose kinetic trapping of Li at commercially relevant current densities.

**Acknowledgements**

The authors appreciate the support from Andreas Sturm, Uwe Tezins and Christian Bross for maintaining the APT and FIB facilities. We also acknowledge Benjamin Breitbach for the kind support on the XRD measurements. T. M. S. and B.G. are grateful for the financial support from German Research Foundation (DFG) through the award of Leibniz Prize 2020 from B.G. T.M.S. gratefully acknowledges the financial support of the Walter Benjamin Program of the German Research Foundation (DFG) (Project No. 551061178). This work was financed, in part, by the São Paulo Research Foundation (FAPESP), Brazil, under Process Number 2025/04646-0. C.J. acknowledges the National Research Foundation of Korea (NRF) grant funded by the Korean government (MSIT) (RS-2024-00359650). P.Y. is grateful for financial support from the Alexander von Humboldt Foundation. Y.J. would like to thank the Horizon Europe under the grant agreement No. 101192848.

**Methods**

### 1.1. Materials synthesis

Li-71.1Mg (at.%) ingots with 200 g were cast under argon atmosphere by mixing the high-purity metals, followed by solidification in Cu molds. After ageing at room temperature for approximately 14 years, the specimens were cut in disk-shape (Ø = 10 mm, thickness = 200 μm) by electric discharge machining. The chemical composition was confirmed by atomic absorption spectroscopy. The damaged surface was removed by dry grinding with #1200 SiC paper until a shiny surface was obtained in a dry room before cell assembly.

. Additional samples of Li-10 Mg (at %) and Li-10 Mg (at %) were prepared by mixing the pure metals Li (Sigma-Aldrich, 99% purity) and Mg (Sigma-Aldrich, 99% purity) in a furnace at 400°C for 4 hours and at 650°C for 6 hours, respectively, in order to obtain a homogeneous composition.

A molybdenum (Mo) crucible was employed to suppress undesired reactions of lithium at elevated temperatures (>450 °C). During the casting process, lithium foil was first pre-melted at 200 °C, after which magnesium (Mg) turnings were gradually added into the molten lithium to promote a more efficient solid–liquid reaction. The melt was stirred at 30 min intervals to enhance the reaction and improve compositional homogeneity. These alloys were pressed (hydraulic press) to form sheets of the respective alloy and used for X-ray diffraction (XRD) and differential scanning calorimetry (DSC).

### 1.2. Cell assembly and electrochemical measurement

2032 coin cells were assembled with a Li counter electrode from Sigma-Aldrich (99.9% purity). Two glass fiber separators (Whatman, 675 μm thickness) were used between the electrodes, and 150 µL of 1 M $LiPF_6$ in EC/DMC (50/50 wt/wt) electrolyte was used. After leaving the sample under open circuit voltage (OCV) for approximately 12 h, galvanostatic lithiation experiments under currents of 20, 25 and 200 µA/cm$^2$ were performed until a capacity of 2.5 – 6 mAh/cm$^2$ was obtained. The temperature was kept constant at 25°C. After discharge, the cell was disassembled within 10 min, in an Ar-filled glovebox (<0.8 ppm $O_2$ and <0.07 ppm humidity). Samples were analyzed either immediately after cell cycling or up to 21 days after. In case of immediate analysis, the disassembled cells were plunged in liquid nitrogen to freeze the microstructure.

### 1.3. Focused ion beam and atom probe tomography

The anode was mounted onto a cryogenic sample holder for APT, and transferred into a high vacuum suitcase (<$10^{-6}$ mbar). The suitcase with the sample inside was cooled down to cryogenic temperatures by filling a dedicated container with liquid nitrogen[1] to cease any Li redistribution for samples analyzed immediately after cell cycling. Micrometric bars of the material were lifted-out using a scanning electron microscope with a Ga-focused ion beam (SEM-FIB), following the procedure in Ref.[2], at cryogenic temperatures (-190°C) from the top surface until 50 µm depth for samples analyzed 1-21 days after cycling. The lift-out procedure for samples analyzed immediately after cycling was performed at cryogenic temperatures (-190°C)[2] to limit Li migration triggered by ion beam exposure.

The sharpening of specimens into needles suitable for APT was performed under cryogenic temperature (-190°C). After milling, a final cleaning step with a circular milling pattern was performed at 5 kV/7 pA for 30 s to remove the residual damaged layer at the surface. The specimens were then transferred to the atom probe through the same suitcase pre-cooled to -196°C. For the characterization of the initial Mg-Li alloy, i.e. before battery operation, room temperature lift-out[3], milling and air transfer were used.

APT measurements were obtained on a Cameca local electrode atom probe LEAP 5000XS, with a straight flight path, in both voltage and laser pulsing modes. In both cases, the base temperature was 50 K, the target detection rate of 1 – 2 ions per 200 pulses on average, and the pulsing rate was 125 – 200 kHz. In laser-pulsing mode, the laser pulse energy was 10 – 45 pJ. In voltage-pulsing mode, the pulse fraction was 30%.

Data reconstruction and analysis were performed using the commercial software AP Suite 6.3 following the voltage reconstruction. The radial distribution function (RDF) and the 1$^{st}$ nearest neighbor distribution (NND) for Li ions were computed in each dataset. In the analysis, since we only detect Mg, Li and some minor amounts of O, we consider the sample as a binary Li-Mg alloy.

Quantification of spinodal decompositions by APT can be done by using proximity histograms[4] , but these are calculated based on a user-defined isosurface, hence the obtained results may differ between researchers. Thus, in order to calculate the compositions of the Li-poor and Li-rich region without any user defined parameter, we use the NND and RDF in the distribution of isolated atoms (DIAM)[5] and pair correlation function (PCF)[6] methods, respectively.

First, the NND was used to calculate the matrix composition using the distribution of isolated atoms (DIAM)[5] for the determination of the matrix composition. This method is based on the calculation of the relative fraction of isolated solutes as a function of distance (d) within the APT reconstruction. For a given solute, the average

NN distance within precipitates is shorter than in the matrix, in such a way that the contribution from both can be separated[7].

Second, the RDF data was used as input to calculate the pair correlation function (PCF)[6], which can be advantageously used to characterize spinodal decompositions[8–10]. The pair correlation, γ(r), as function of distance was modeled based on the following equation, which is derived from classical small-angle scattering models[11–13]:

$$\gamma(r) = < \Delta C^2 > e^{-\frac{r}{\xi}} \frac{sin\frac{2\pi}{\lambda}r}{\frac{2\pi}{\lambda}r}$$

Where $\langle \Delta C^2 \rangle$ is the mean square fluctuation, λ and ξ are the wavelength and correlation length, respectively, in nm. The correlation length can be seen as the typical size of the features, while the wavelength is the periodic modulation between Li-rich and Li-poor regions. We consider that Li-rich regions have a uniform composition in a matrix of uniform composition separated by a sharp interface. The matrix composition, i.e. Li-poor regions, can be quantified by the DIAM method. The mean square fluctuation can be then used to calculate the composition of the Li-rich region, considering the spinodal decomposition as a 1D sinusoidal profile[14]:

$$C_{max} = \sqrt{2\langle \Delta C^2 \rangle} + C_{min}$$

Where $C_{max}$ is the Li content in the Li-rich region, $C_{min}$ is the Li concentration in the Li-poor region (matrix).

### 1.4. Scanning electron microscopy and X-ray diffraction

A cross section of the specimen was prepared by cutting the sample with a scissor inside the Ar glovebox, followed by air transferred into the SEM-FIB. The cross section was milled at -190°C to remove the deformation layer from the previous step. An acceleration voltage of 30 kV was used, with the current decreasing step-wise until 9.3 nA. The specimen was air transferred to a Zeiss Merlin SEM within less than 2 minutes. Backscattered electron images were acquired under an acceleration voltage of 15 kV, with a current of 5 pA to minimize electron beam damage.

X-ray diffraction was performed on the top part of the as received sample under air. A Cu radiation source was employed (λ = 0.154059 nm) in the conventional ϴ-2ϴ configuration, scanning angles between 20˚ and 130°, with a step size of 0.03° while spinning the sample around the surface normal vector to reduce texture effects.

### 1.5. Transmission electron microscopy

The TEM cross-sectional lamella specimen was prepared using a Thermo Fisher Scios 2 dual-beam focused ion beam (FIB)–SEM system. Pt was deposited on the region of interest to prevent Ga implantation and its associated artificial effects. Additional cleaning milling was performed sequentially at 5 and 2 kV to remove the amorphized surface layer caused by Ga-ion damage. Transmission electron microscopy (TEM) investigations were conducted using an image-aberration-corrected Titan Themis 80–300 microscope (Thermo Fisher Scientific) operated at an acceleration voltage of 300 kV. Electron diffraction pattern was acquired using a 400 nm diameter aperture for selected-area electron diffraction (SAED).

The hcp planes in the SAED patterns were indexed using ICSD database code 104740. To describe the B2 phase, ICSD database code 57596 was modified by adjusting the lattice constant to 0.3486 nm according to the reported reference value[15] for BCC LiMg. From the SAED analysis, the measured d-spacing of 0.207 nm could not be assigned to any hcp planes, whereas it matched with the {111} planes of the B2 phase. The slight discrepancy between the measured value (0.207 nm) and the calculated value (0.201 nm) is considered to originate from the ordering effect in the B2 structure.

### 1.6. Differential scanning calorimetry and Li diffusion distance calculation

Differential scanning calorimetry was performed from room temperature to -140°C, followed by heating back to room temperature under a cooling/heating rate of 2°C/min. In another set of experiments, DSC was performed from room temperature to 500°C with cooling/heating rates of 10°C/min to crack the order-disorder transformation for the pristine Li-50 at.% Mg alloy. In both cases, the heating and cooling cycles were performed 3 times to confirm the consistency of the observed peaks.

The Li diffusivity at a given temperature is calculated based on an Arrhenius equation

$$D = D_0 \exp\left(\frac{-E_a}{RT}\right)$$

Where $D_0$ is the diffusion coefficient at infinite temperature, $E_a$ is the activation energy, $R$ is the universal gas constant and $T$ is the temperature. Considering a Li-20at.% Mg alloy, the diffusion coefficient at 298 K is 6.31 x $10^{-12}$ $cm^2s^{-1}$ and the activation energy is 58.662 kJ/mol[16]. The diffusivity at 133 K (-140°C) is calculated as 1.1 x $10^{-24}$ $cm^2/s$. Under the second Fick's law, considering constant concentration Li source in a 1D diffusion, the diffusion length for Li in a certain amount of time can be calculated as

$$L = 2\sqrt{Dt}$$

Where $t$ is the time.

### 1.7. Calculation of the Li-Mg phase diagram

The Li–Mg system was assessed in this work using the thermodynamic parameters reported by Wang et al.[17] for the liquid, BCC_A2, and HCP_A3 phases as a starting point. To properly describe the ordered BCC_B2 phase observed in this work, the order–disorder formalism[18] was applied, together with a crystallographically consistent four-sublattice thermodynamic model to represent the four distinct crystallographic sites of this phase. The formation energies of the endmember compounds—i.e., configurations in which each sublattice is occupied by a single element—were obtained from density functional theory (DFT) calculations reported by Taylor[19] and from the Open Quantum Materials Database (OQMD)[20]. Finally, the temperature-dependent interaction parameters were optimized to reproduce the experimental data presented in this work. A more detailed description of the assessment is beyond the scope of the present work; additional details and possible refinements may be addressed in future work.

**Conditional spinodal decomposition in Li-Mg anodes for lithium metal batteries**

Leonardo Shoji Aota[a,b], Aubin Leray[c,d], Yuqi Liu[a], Frederic de Geuser[e], Chanwon Jung[a,f], Shyam Katnagallu[a], Tim M. Schwarz[a], Alisson Kwiatkowski da Silva[a,g], Júlio César Pereira dos Santos[g], Eric Marchezini Mazzer[h], Poonam Yadav[a,i], Christoph Freysoldt[a], Frank Stein[a], Yug Joshi[a], Se-Ho Kim[a,j], Dierk Raabe[a], Baptiste Gault[a,k]

## Supplementary Material

### Supplementary Discussions

### 1.1. Pristine Li-71.1 at.% Mg alloy

The X-ray diffractogram in Figure S1b highlights only the presence of α-HCP Mg, although we expect a two-phase microstructure with α-HCP Mg and β-BCC, based on equilibrium thermodynamics[21]. Improper indexation of the β-BCC matrix also takes place in Figure S1a. Yet, we hypothesize two regions by this combination of scanning electron microscopy and X-ray diffraction: a coarse-grained β-BCC matrix with dispersed α-HCP grains. Despite storing the sample for 14 years at room temperature, the identification of three phases, β-BCC matrix, α-HCP grains and 50 at.% Li precipitates within the α-HCP grains already indicate that despite long-term storage the system has not reached equilibrium according to the reported phase diagram.

## 1.2. Enthalpy of mixing and Gibbs free energy of α-HCP and β-BCC phases

The enthalpy of mixing in Figure S2a calculated with the SSOL6 database indicates the decrease or increase of internal energy caused by chemical bonds between atoms in the alloy. A positive enthalpy of mixing reflects the increase in internal energy when Li and Mg atoms are positioned as neighbors, according to the quasi-chemical model; the system then will tend to create clusters of Li-rich and Li-poor (or Mg-rich) regions. A positive enthalpy of mixing is then responsible for the immiscibility of two solid solutions, as observed in classical spinodal decomposition.

In contrast, a negative enthalpy of mixing is caused by a decrease in internal energy when Li and Mg are neighbors. A crystal arrangement where a continuous alternating Li-Mg arrangement is possible is the ordered structure. The ordering may be long- or short-range, depending on the entropy of mixture being low enough to allow long-range ordering. The negative enthalpy of mixing of β-BCC phase in Figure S2a throughout its stability range, based on current databases, indicates a favoring condition for the formation of a long-range ordered B2 structure.

From the free energy curve in Figure S2b, there is no negative curvature, which would be caused by a positive enthalpy of mixing, in the β-BCC phase. Thus, from the current databases, it is unlikely that the β-BCC undergoes spinodal decomposition. We thus argue that the formation of the ordered B2 structure is a pre-condition to trigger the spinodal decomposition reported in this manuscript.

### 1.3. Phase transformations under cryogenic temperatures

The possibility of phase transformations taking place during sample cooling to 50 K (-223°C) was discarded as no measurable phase transitions are detected in cryogenic DSC in Figure S4. Considering a diffusion length of 10 nm required to produce the spinodal fluctuations (similar to the wavelength) and the diffusivity at 133 K (-140°C), we calculate the required time as 2.26 x $10^{11}$ s (2,615,740.7 days). Thus, Li mobility is not sufficient to cause the spinodal decomposition at cryogenic temperatures, considering the time required for the sample preparation at -190°C in the cryo-FIB and the measurement time in the APT.

In addition, the equilibrium phase diagram indicates a martensitic BCC→HCP phase transformation at around -196°C[22] for Li-rich alloys. However, the alternating Li-poor and Li-rich are unlikely to be caused by martensitic transformation, since such transformation is displacive[23] and does not involve long-range diffusional processes required to form such patterned chemical fluctuations. The laser pulses may allow Li movement during APT measurements[24]; however, no sign of Li migration was observed, with a stable voltage curve and low background throughout all measurements (Figure S10). Thus, we are confident that spinodal fluctuations are not resulting from measurement artefacts and it indeed takes place in the BCC phase in the Li-Mg system.

### 1.4. On the spinodal decomposition

Most observations of spinodal in solids were taken in thermomechanically processed alloys, especially when water quenching (fast cooling) a homogenized alloy, followed by a low temperature annealing to start the decomposition[25,26] i.e., a closed system. In the present case, the spinodal is triggered by a constant pressure and temperature process, where an external electrochemical potential drives continuous Li ingress in the Mg-Li alloy, reaching the miscibility gap in this system.

The spinodal decomposition has no energy barrier[27]. However, the elastic energy between Li-rich and Li-poor regions may alter the stability criterion[28]. When the elastic strain energy is too high, the spinodal decomposition can be hindered, resulting in a nucleation and growth-mediated precipitation of non-spherical particles[29]. These fluctuations are modelled as sinusoidal functions along directions with lower elastic energy, such as <100> or <111> in the BCC structure[30]. They systematically decompose with the shortest wavelengths, and hence most rapidly, since it requires a shorter Li diffusion distance. However, other directions with slightly higher elastic energies can also trigger such fluctuations, especially within the center part of the miscibility gap, where the stability of such structure is higher[30].

During the early stages of spinodal decomposition, the dominant plane grows exponentially, resulting in purely sinusoidal fluctuations in <100> or <111> planes[30]. In the initial stage of decomposition, the chemical composition amplitude increases with time, while the wavelength is constant. In the second stage, both amplitude and wavelength increase, indicating a coarsening of the structure driven by minimization of the interfacial free energy[31]. We expect the development of phase separation under a certain average chemical composition to be more advanced under lower current densities, since more time will be available for phase separation until such Li content is reached. It allows more time for the Li-rich and Li-poor regions to reach the compositions given by the spinodal curve in the phase diagram. By increasing the current density, the evolution of spinodal decomposition for the same average Li content is lowered until a threshold value is reached. Further increase of current density will not affect the spinodal structure due to Li diffusivity limitation in β-BCC.

Grain boundaries can act as fast diffusion paths, while also attracting Li in thermodynamic equilibrium. The grain boundary is enriched in Li, Figures 3f-g and Figures S3a-d, and its spinodal fluctuations exhibit a longer wavelength (Figure S2b). The grain boundary is already in the second stage of spinodal decomposition, while the adjacent grains are in the first stage, as suggested by the higher wavelength for the former (orange icon in Figure S2b).

### 1.5. Field evaporation of Li-rich alloys (DFT calculations)

APT analysis of pure Li is known to be challenging[32]. Yet, we observe a continuous voltage curve during atom probe measurement and a stable background below 10 ppm/ns in Figure S11, both indicating a stable measurement[24], without field evaporation artefacts. We performed density functional theory calculations to understand what the underlying reason for the stability of our measurement is, despite reaching local Li contents above 95 at.% Li in many datasets.

We constructed atomically defined metallic surfaces using a custom-built slab generator capable of producing high-Miller-index orientations[33]. The tool enables precise specification of terrace width, step geometry, and kink direction, following the methodology originally proposed by Van Hove and Somorjai[34]. All models and density functional theory (DFT) calculations were managed within the pyiron environment[35], which provides a unified interface for structure generation, job submission, and post-processing.

A body-centered cubic Li (110) slab containing 65 atoms was chosen as the representative system. An external electrostatic field was introduced asymmetrically across the slab through the generalized dipole correction scheme implemented in S/PHI/nX[36,37], ensuring a well-defined potential drop on one surface. Electronic structure calculations employed the projector augmented-wave (PAW) pseudopotentials from the VASP library in combination with the Perdew–Burke–Ernzerhof (PBE) exchange–correlation functional[38]. Convergence criteria were established to guarantee total-energy accuracy within $10^{-3}$ eV, achieved using a plane-wave cutoff of 550 eV and a k-point sampling of 6 × 6 × 1 for the (110) orientation, with equivalent k-density maintained for higher-index surfaces such as (952).

Three different starting configurations were chosen for relaxation calculations at an electric field of 1.5 V/Å. These three structures differ in the position of one substitutional Mg atom, viz., Mg in the terrace, Mg in the step and finally Mg below the step. The starting configurations are plotted in the top row in Figure S12 (a, b, c). The final relaxation patterns at this field are shown in the bottom row of Figure S12, corresponding to each configuration. Only the top layer atoms (Li and Mg) are plotted for clarity and are colored according to the amount of relaxation in 'z' direction from the starting structure. In the case where Mg is in the step next to Li kink atom, spontaneous evaporation of Li is triggered, as seen in Figure S12b. For the rest, no evaporation is observed. The earlier DFT calculations for pure Li[39], suggests a ~0.05 eV evaporation barrier at 1.5 V/ Å for Li kink atom from the same surface. From these preliminary calculations, a conclusion can be drawn that Mg has an effect of decreasing the evaporation barrier of Li, especially when it is a direct neighbor in the step. The evaporating Li also rolls over the step Mg atom. The reduction in evaporation barrier of Li, also suggests that a uniform evaporation should be observed of Li, against the formation of an adatom gas phase as predicted in ref[39] for pure Li leading to an inhomogeneous evaporation. Therefore, we support the reliability of our Li quantification from atom probe tomography, without artefacts arising from inhomogeneous field evaporation of Li atoms due to the presence of Mg.

### 1.6. Origin of a two-phase microstructure from a spinodal precursor

A two-phase microstructure, i.e., characteristic from nucleation and growth, can also be promoted by a spinodal precursor. These transformation stages were already predicted by Cahn, where sinusoidal compositional fluctuation waves from higher orders, i.e. <210> and <300>, interact with the initially dominant <100> wave[40], when a critical amplitude is reached. As a result, odd waves (h+k+l = odd) distort the harmonic waves, flattening the composition extremes between Li-rich and Li-poor regions and sharpening the composition gradient at the interface. It slows down the amplitude growth of the original wave, gradually giving rise to two distinct chemically homogeneous phases separated by a sharp interface[40], i.e. a two-phase microstructure.

### 1.7. Kinetic effect on the spinodal decomposition

For compositions that are unstable in a single phase in the phase diagram[30], such as near the limit of metastability[28], spinodal decomposition may be kinetically hindered and superseded by a nucleation and growth mechanism. Two scenarios could be envisaged. First, if the Li diffusivity in B2 phase is insufficient to overcome the spinodal decomposition (sweep the whole unstable composition range), the latter will take place regardless of the current density. In this case, higher current densities will favor Li deposition at the surface of the Li-Mg foil, with simultaneous diffusion into the Li-Mg alloy, creating a large chemical composition gradient within the alloy. Second, if the Li diffusivity in the B2 phase is high enough, spinodal decomposition may be avoided at high current densities. The fast Li ingress and redistribution may shift the local chemical composition outside the limit of metastability before the spinodal decomposition is started, resulting in a phase transformation via nucleation and growth mechanism or the direct formation of a homogeneous Li-rich solid solution.

## 2. Supplementary Figures

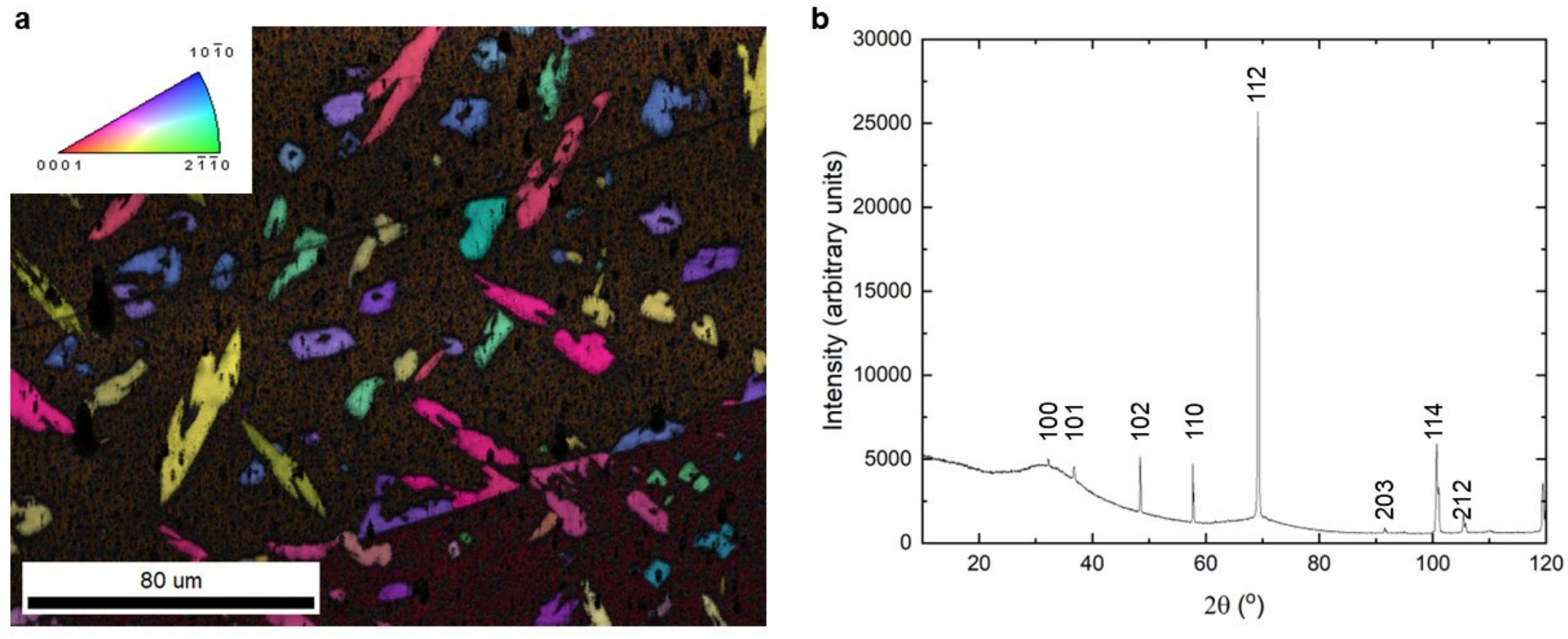

**Figure S1 –** (a) Inverse pole figure map with directions perpendicular to the top surface of the anode, obtained from electron backscatter diffraction, highlighting the presence of finer α-HCP grains within likely coarser β-BCC grains. The improper indexation of the matrix is likely caused by Li removal during metallographic preparation with alcohol-based suspensions and solutions. (b) X-ray diffraction profile from the pristine Li-71.1 Mg (at.%) alloy, acquired with Cu X-ray source, indicating only α-HCP peaks.

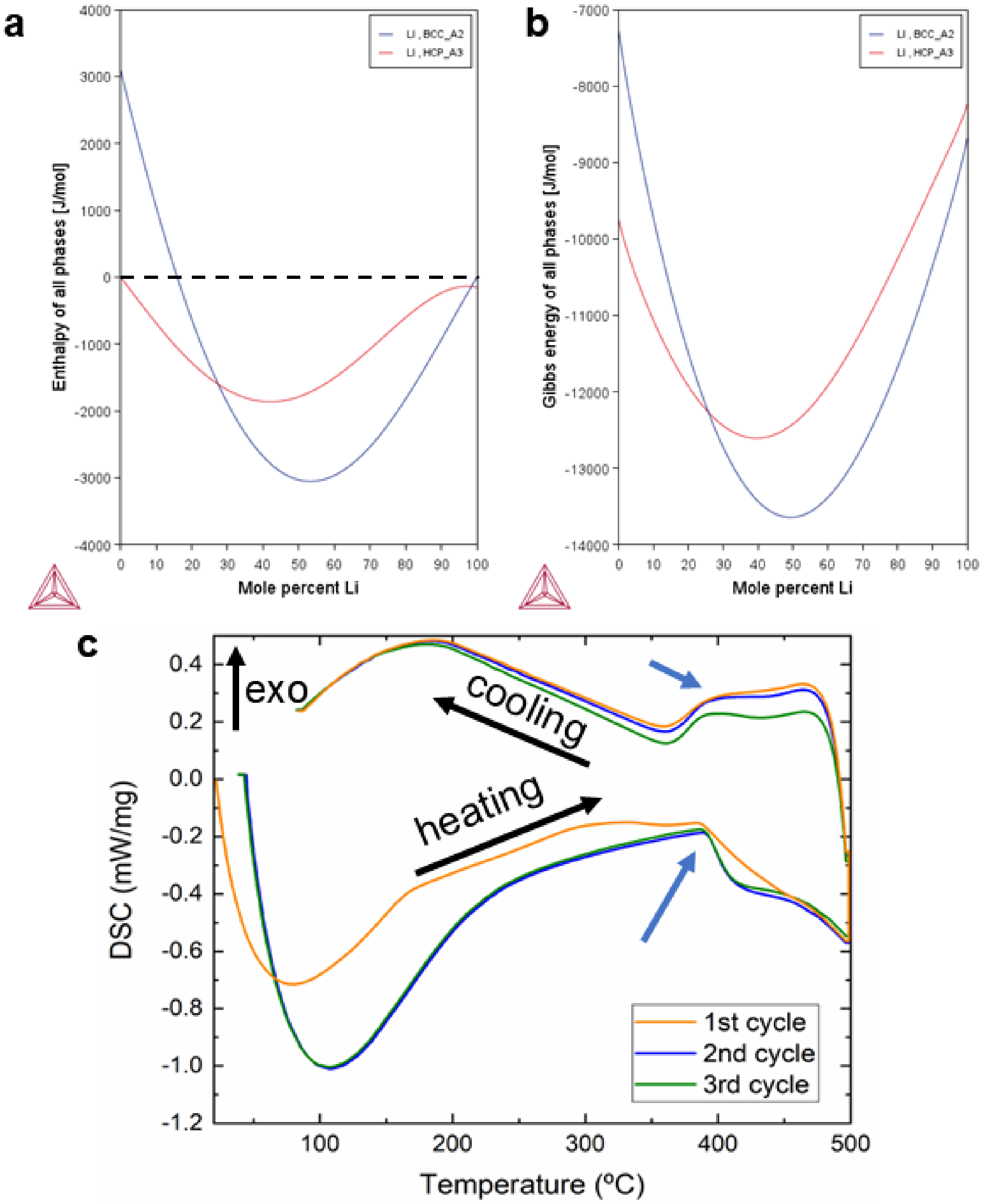

**Figure S2 –** (a) Enthalpy of mixing and (b) Gibbs free energy as function of the Li concentration (atomic/mole percent) in the Li-Mg system for both β-BCC and α-HCP phases at 25°C, calculated with the current SSOL6 database. The enthalpy of mixing of the β-BCC phase in (a) is negative for most of the composition range, i.e., in the composition range where the β-BCC phase is thermodynamically stable. It reflects the decrease in internal energy when Li and Mg are neighbors, according to the quasi-chemical model, which is translated into a tendency of chemical ordering in this phase. As a result of the negative enthalpy of mixing, the free energy curve for the β-BCC phase has only one positive curvature (upwards), contrasting to a negative curvature present in alloys with spinodal decomposition. (c) Differential scanning calorimetry (DSC) curves obtained at heating and cooling rates of 10 K/min for the as-cast Li-50 Mg (at.%) alloy, with a first-order transition at 390 ± 3°C indicated by blue arrows. The direction of the exothermic reaction is indicated by a black arrow and the writing

“exo”. The measurement was carried 3 times to confirm the reversibility of this transformation.

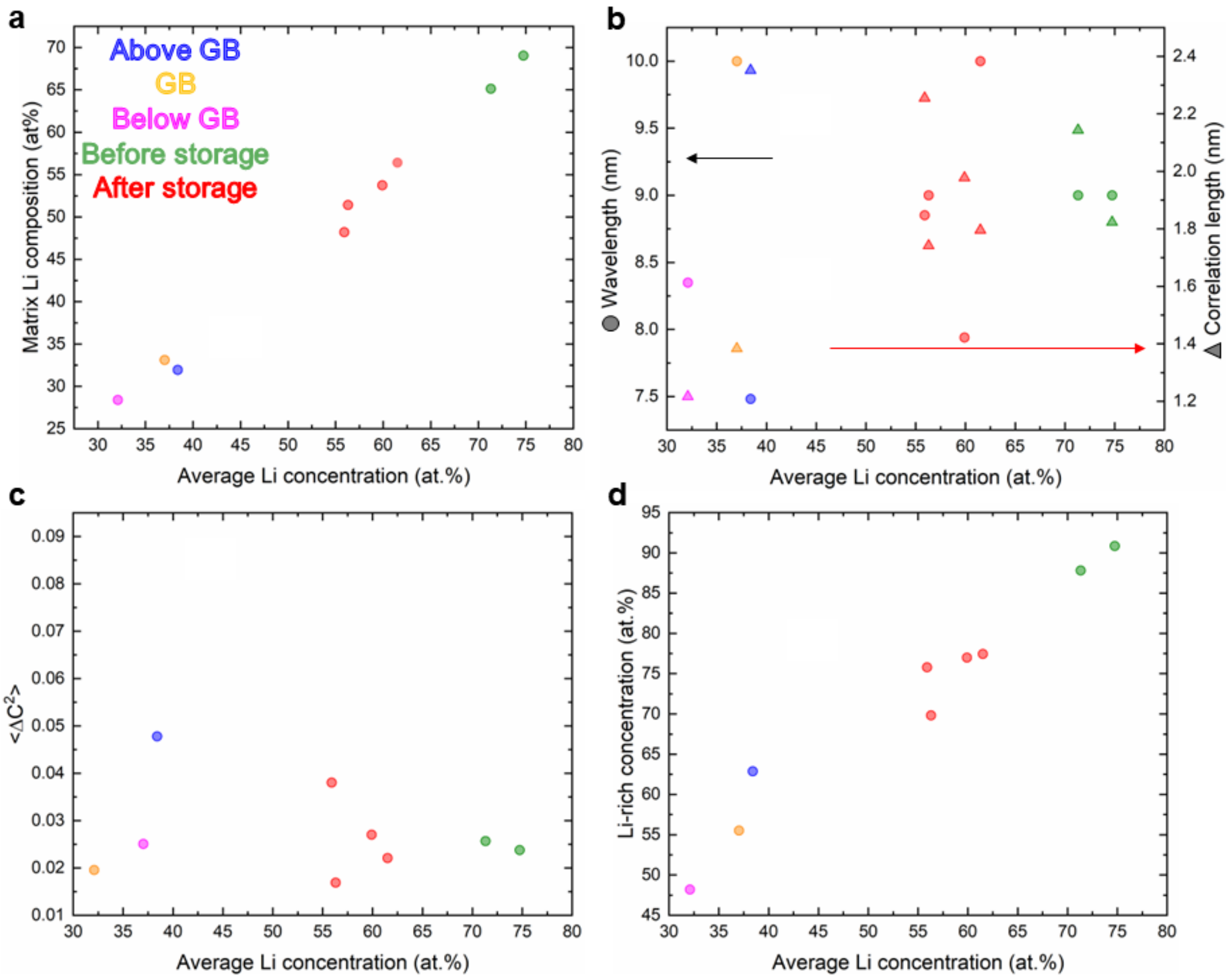

**Figure S3 –** DIAM and PCF analysis of the samples stored at room temperature for 3 weeks. Data from the sample lithiated at 25 μA/cm$^2$ and analyzed directly after lithiation is shown in green for comparison. (a) Matrix Li composition, (b) wavelength and correlation length of the spinodal fluctuations, (c) chemical amplitude and (d) Li concentration in Li-rich regions as function of average Li concentration. Red icons indicate all data obtained after storage. At 50 μm depth, three different regions are distinguished and shown with different icons due to distinct spinodal structures: above GB, GB and below GB, where GB stands for grain boundary.

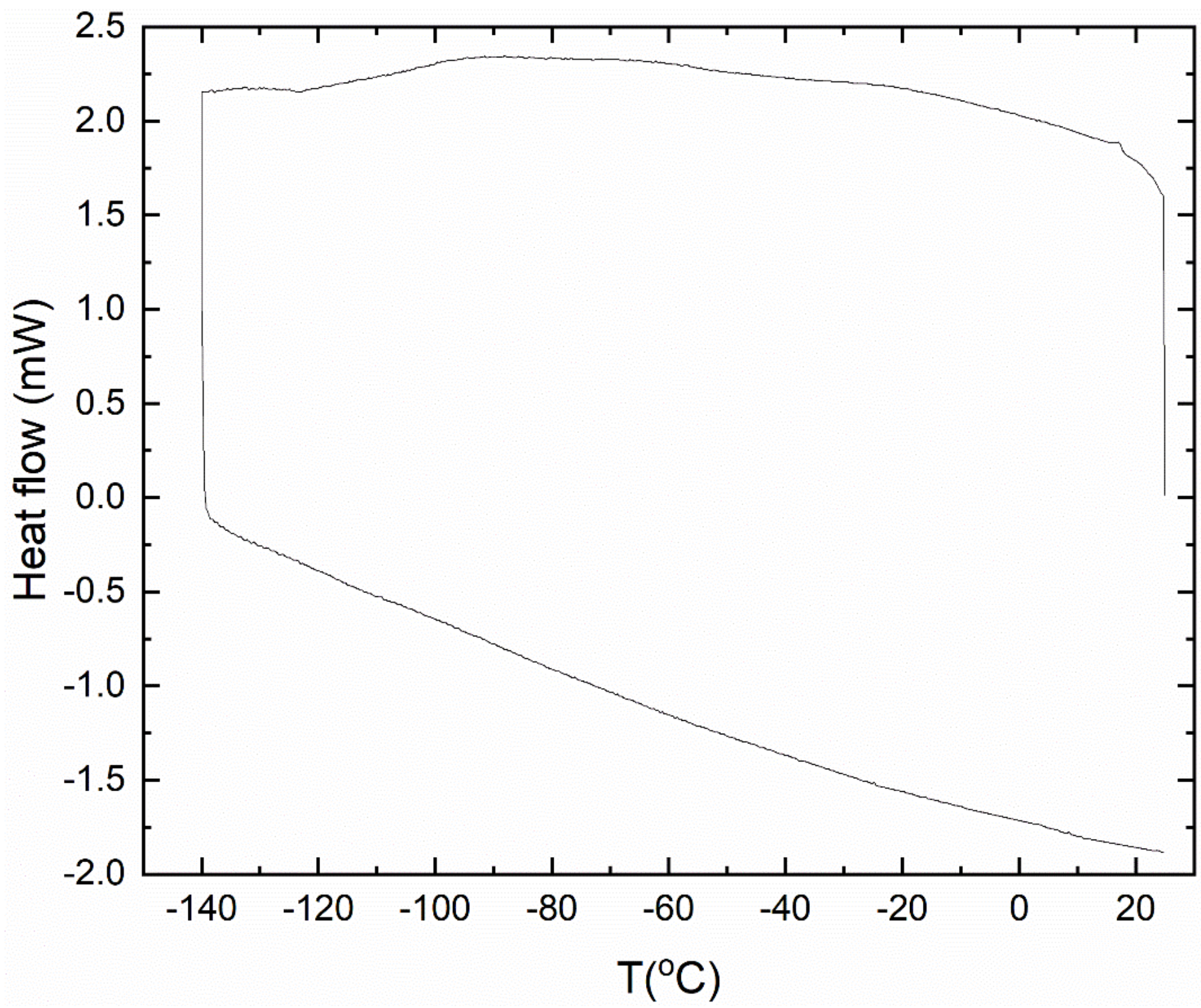

**Figure S4 –** Cryogenic DSC curve of the galvanostatically lithiated sample (25 μA/$cm^2$) at 2 K/min heating and cooling rates. No peaks indicating first or second order transitions are observed upon cooling.

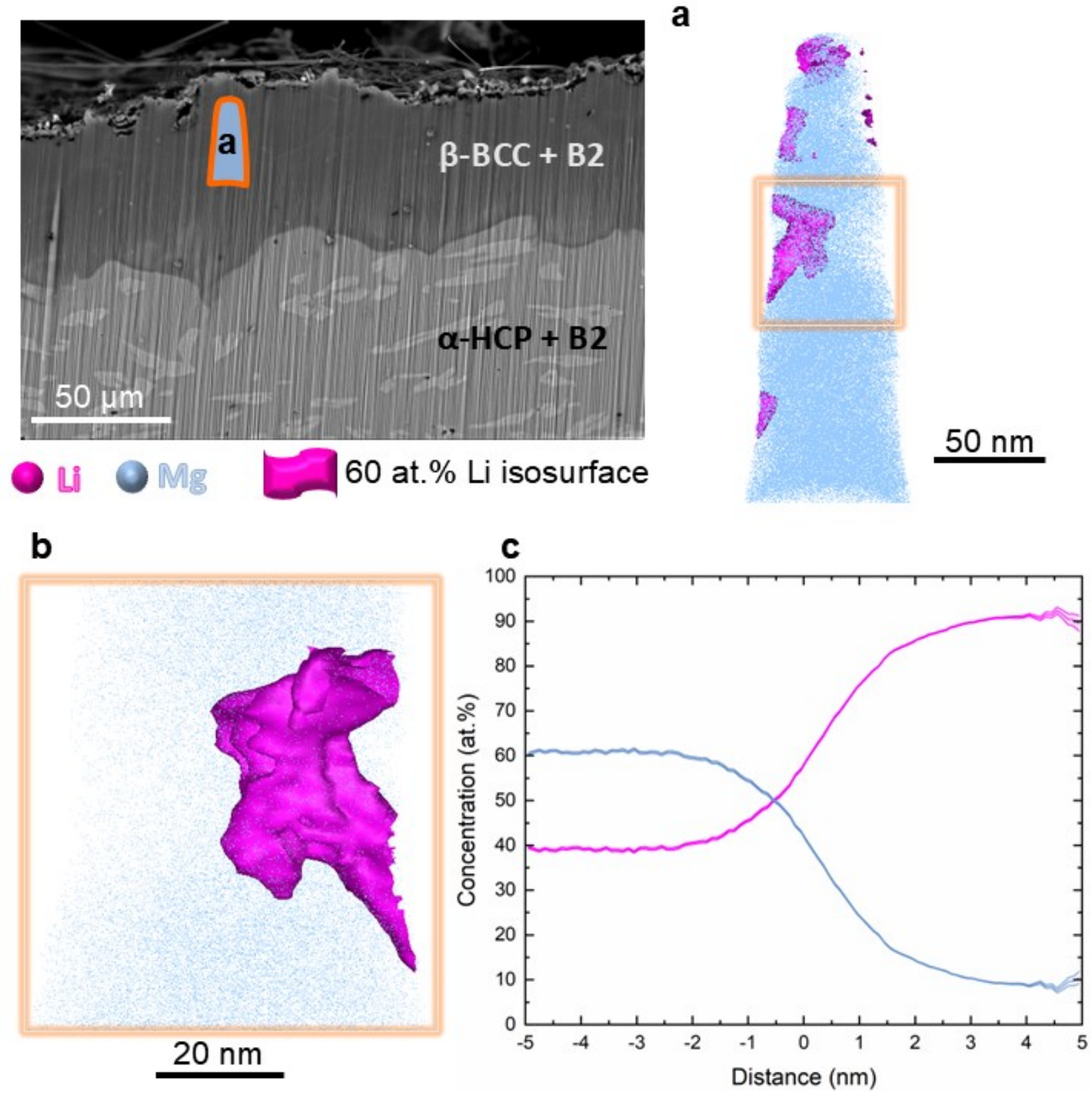

**Figure S5 –** Backscattered electron image, where two distinct layers are observed in the Li-71.1 at.% Mg alloy, lithiated at 20 μA/$cm^2$ until 6 mAh/$cm^2$ areal

capacity and stored for 3 weeks in Ar glovebox. The top layer is enriched in Li, with β-BCC + B2 phases, whereas the bottom layer has both α-HCP + B2 phases. (a) Atom probe reconstruction from the top layer, similar to Figures 4a-c, although with a distinct microstructure. Instead of the spinodal fluctuations, this atom probe reconstruction depicts a two-phase microstructure, with a Li-rich precipitate enlarged in (b) with a 60 at.% Li concentration isosurface. (c) The Li concentration in the Li-rich precipitate and the matrix is around 90 and 40 at.% Li, respectively. Note the homogeneous chemical composition of both regions, which characterizes it as a two-phase microstructure, without Li chemical fluctuations observed in spinodal decomposition.

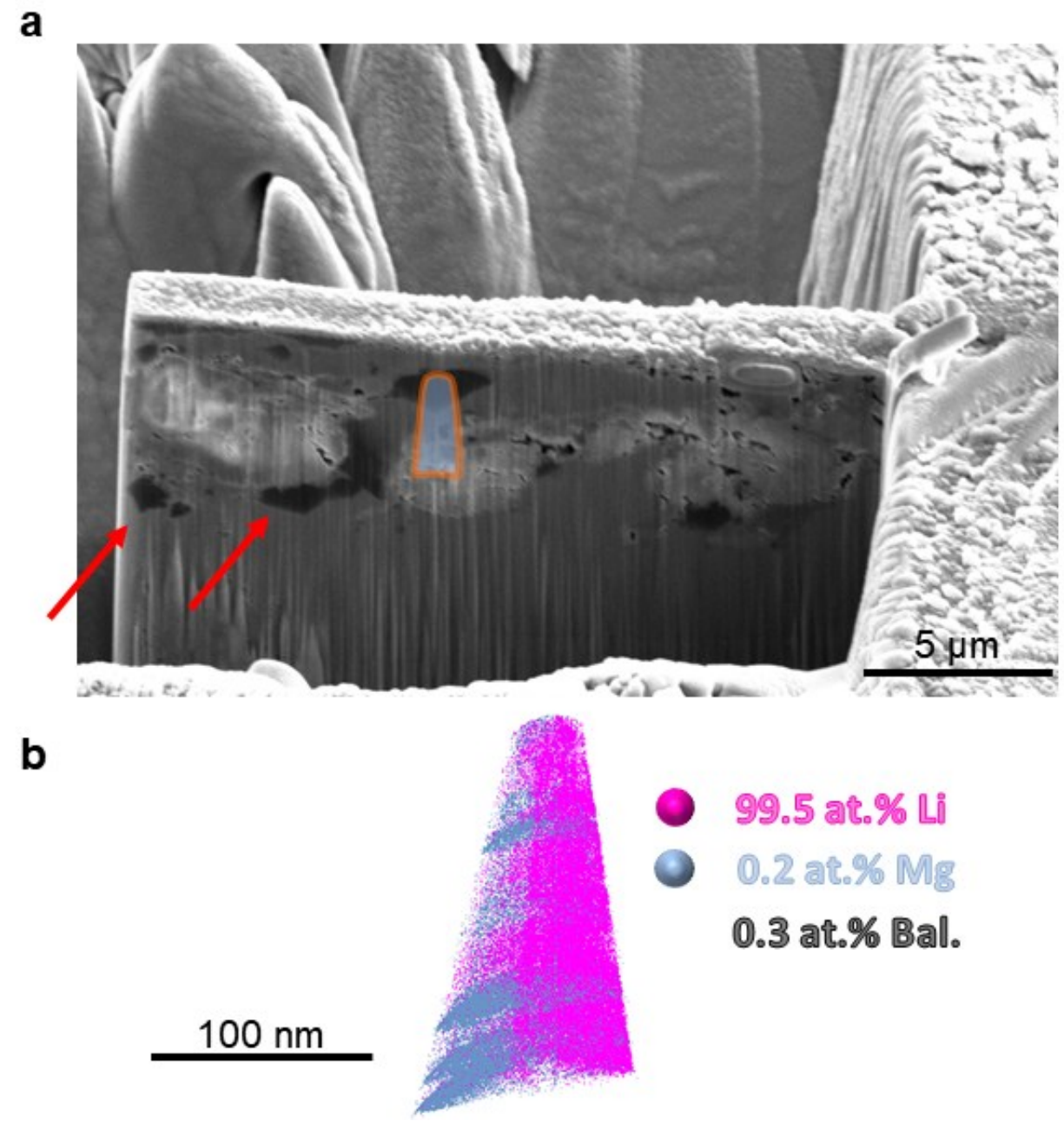

**Figure S6 –** Li-71.1 at.% Mg anode after lithiation at 20 µA/cm$^2$ until 6 mAh/cm$^2$ and room temperature storage for 3 weeks. (a) Scanning electron microscope image taken with backscattered electrons in the focused ion beam under cryogenic temperature (-190°C). The atom probe tomography tip (b) was taken from a "black region", as indicated by the blue tip in (a). Such "black regions" correspond to precipitates with >99.5 at.% Li. Other Li-rich precipitates, i.e. black regions, are indicated by red arrows in (a).

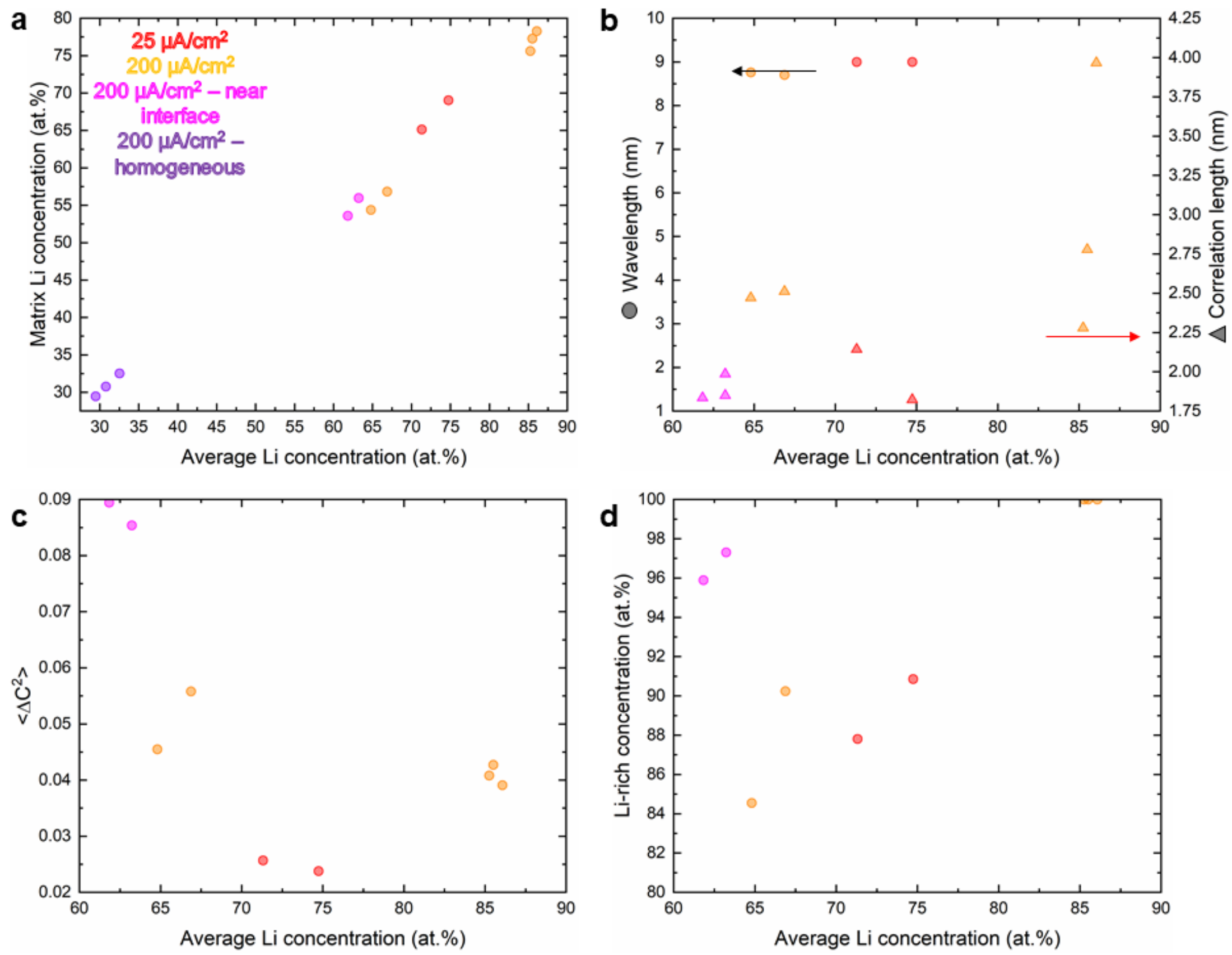

**Figure S7 –** (a) Matrix composition calculated by the DIAM method. (B) Correlation length (ξ, in triangles) and wavelength (λ, in circles). (c) Mean square fluctuation $\langle \Delta C^2 \rangle$, or the chemical amplitude between Li-rich and Li-poor regions, calculated based on the PCF method. (d) The concentration of Li-rich regions based on the parameters obtained from both DIAM and PCF methods. Atom probe tomography data was used as input for DIAM and PCF methods.

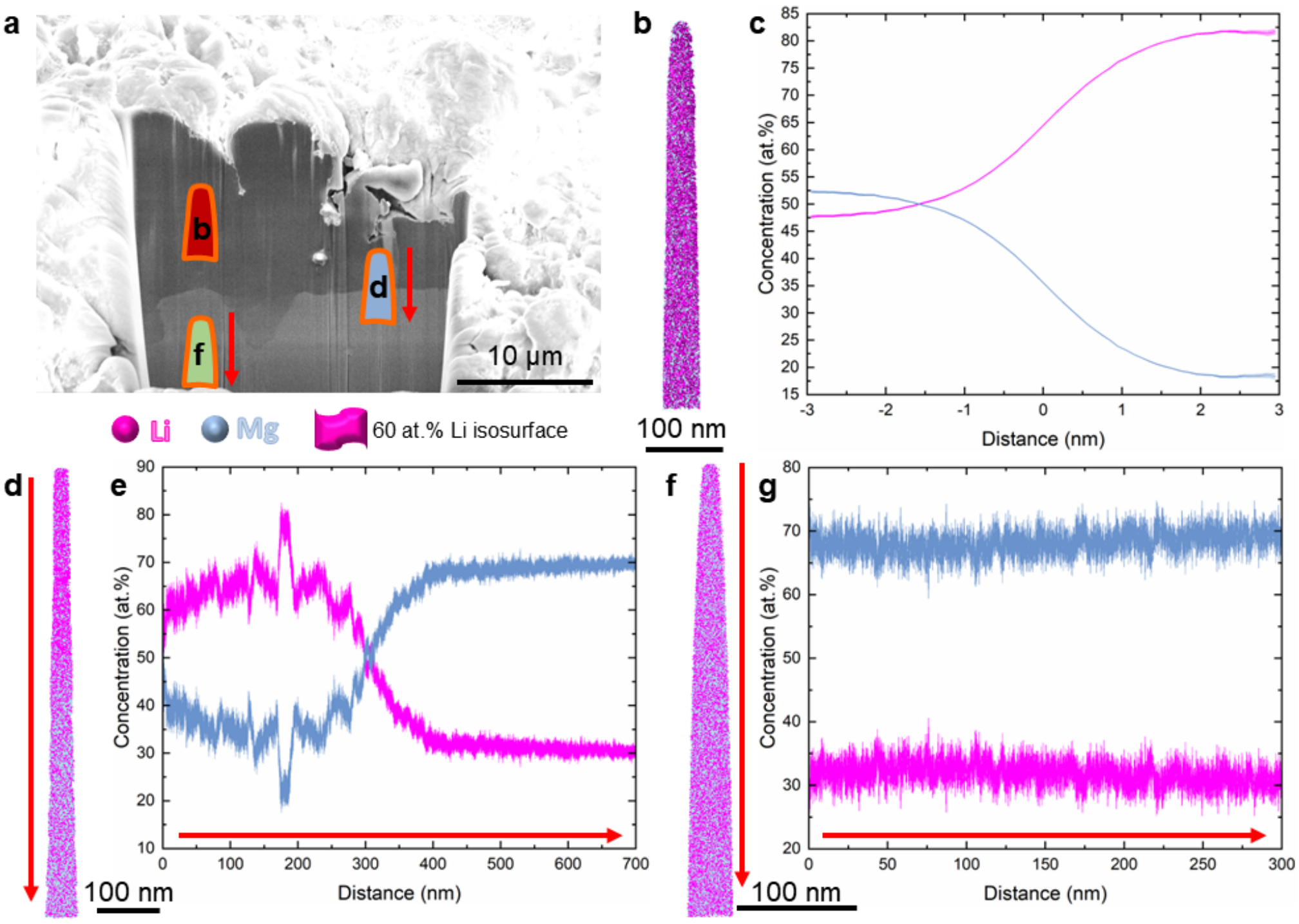

**Figure S8 –** Atom probe tomography data from the sample immediately analyzed after lithiation at 200 μA/cm$^2$ in Figures 4e-f. (a) Cross section of the sample, with two visible layers. The locations of the atom probe tomography reconstructions from where the proximity histograms and 1D chemical concentration profiles were taken are schematically shown. (b) Atom probe reconstruction from the top dark gray region and its corresponding (c) proximity histogram, calculated from a 60 at.% Li isoconcentration surface. (d) Atom probe reconstruction across the interface between dark and light gray layers, with the red arrow indicating the direction of the (e) 1D chemical concentration profile. (f) Atom probe reconstruction from the bottom light gray region, indicating a Li concentration around 33 at.% Li, near the composition from the pristine alloy.

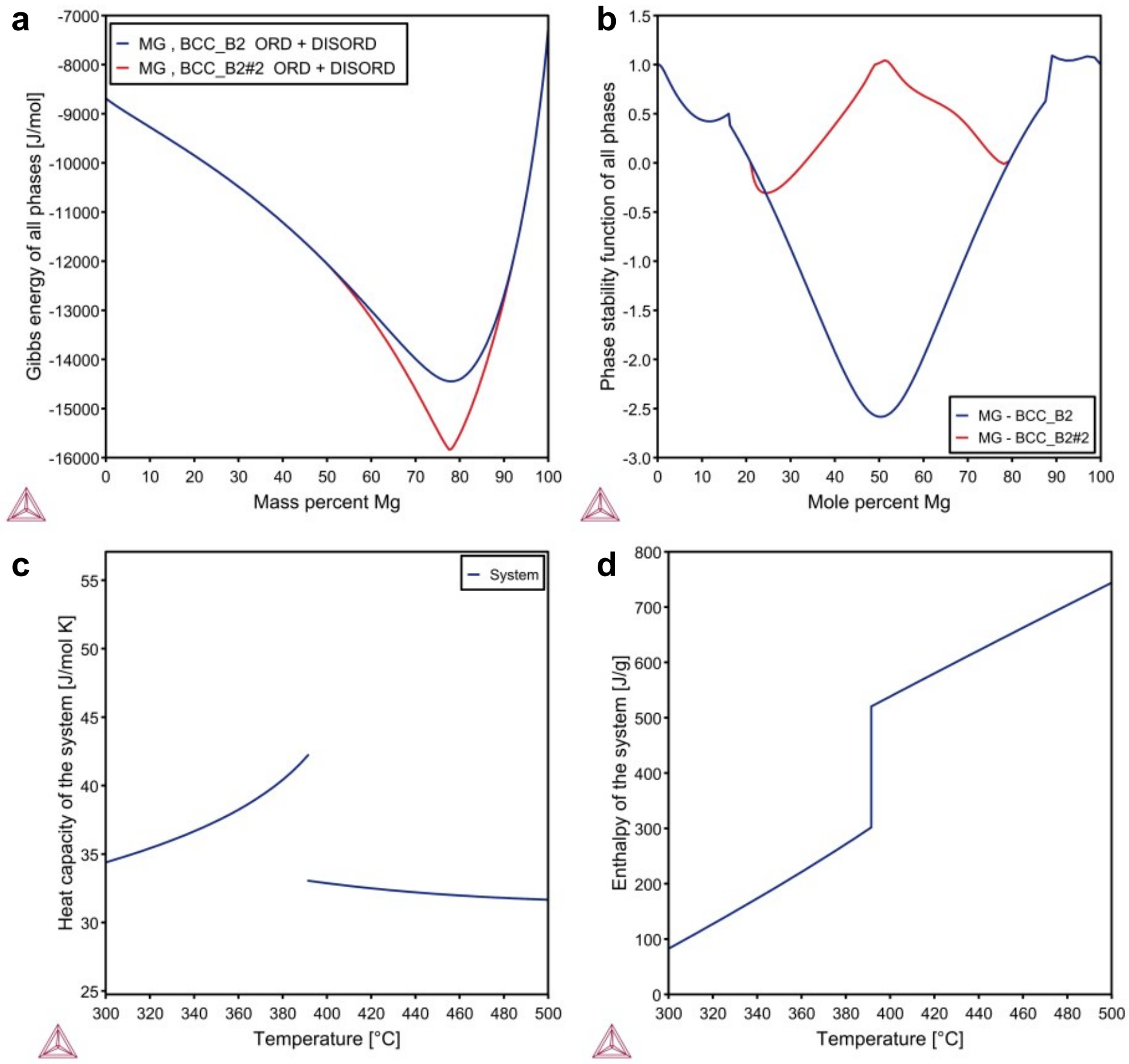

**Figure S9 –** Thermodynamic parameters for the calculated Li-Mg phase diagram, considering a first-order phase transition between B2 and β-BCC phases. (a) Gibbs free energy from BCC_B2 phase. (b) Phase stability function of BCC_B2 phase, indicating the occurrence of a miscibility gap. (c) Heat capacity at constant pressure ($C_p$) and (d) enthalpy for the β-BCC phase. The heat capacity tends to infinite at the transition point for a Li-50 at.% Mg alloy, whereas the enthalpy exhibits a discontinuity at the same point (≈ 390°C, a similar temperature for the 1$^{st}$ order transition measured by DSC).

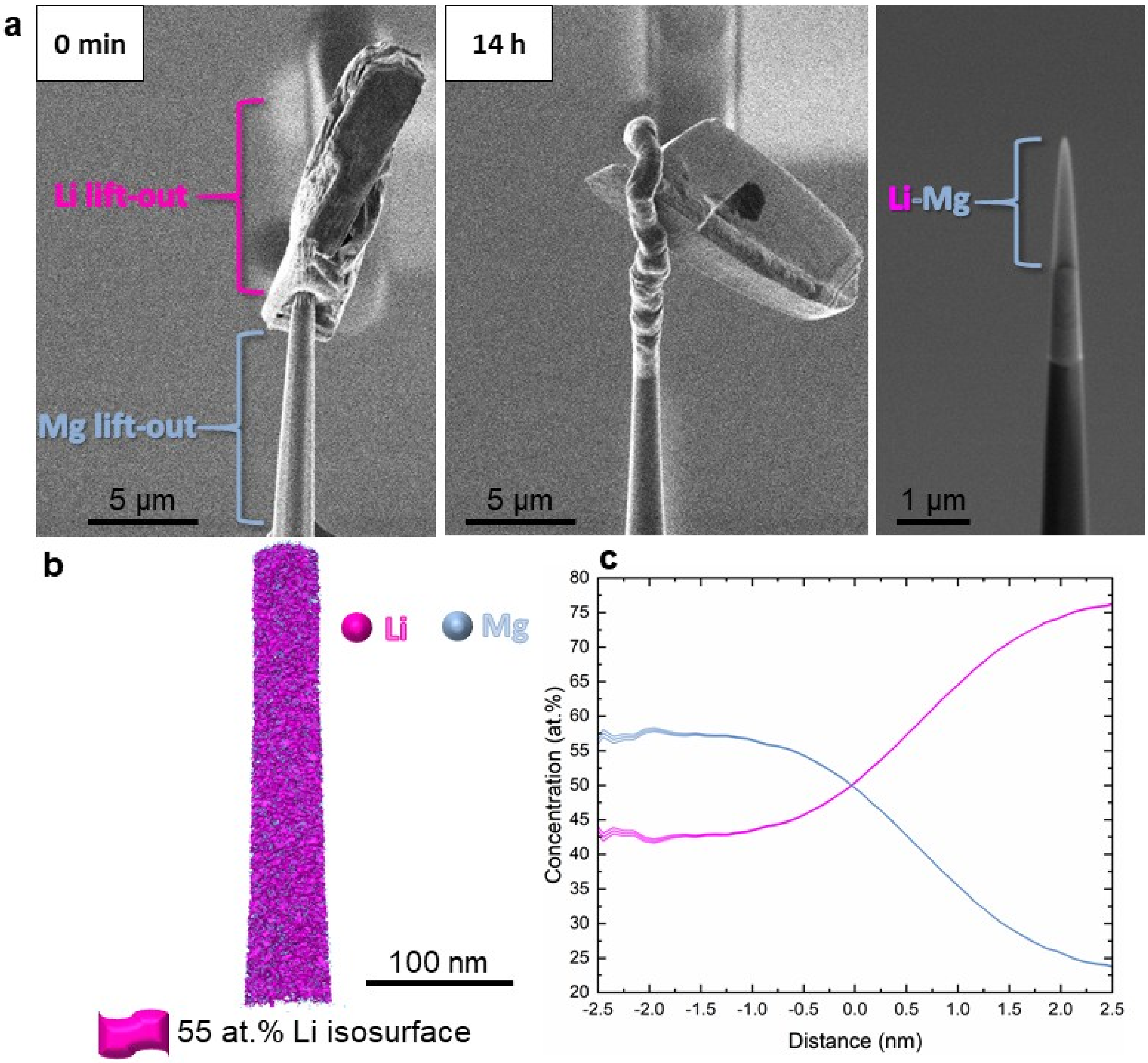

**Figure S10 –** (a) Mg-Li micro-diffusion couple produced by touching a Li lift-out with a Mg pillar. After 14 h reaction at 30°C inside the SEM-FIB, the micro-diffusion couple is cooled down to -190°C for sharpening into a tip shape. (b) Atom probe reconstruction after high vacuum (<$10^{-6}$ Pa) sample transfer from SEM-FIB to the APT at cryogenic temperature (-196°C). Spinodal fluctuations are observed, as indicated by the 55 at.% Li isosurface. (c) Proximity histogram taken from the 55 at.% Li, indicating Li-poor regions containing around 43 at.% Li and Li-rich regions reaching 76 at.% Li.

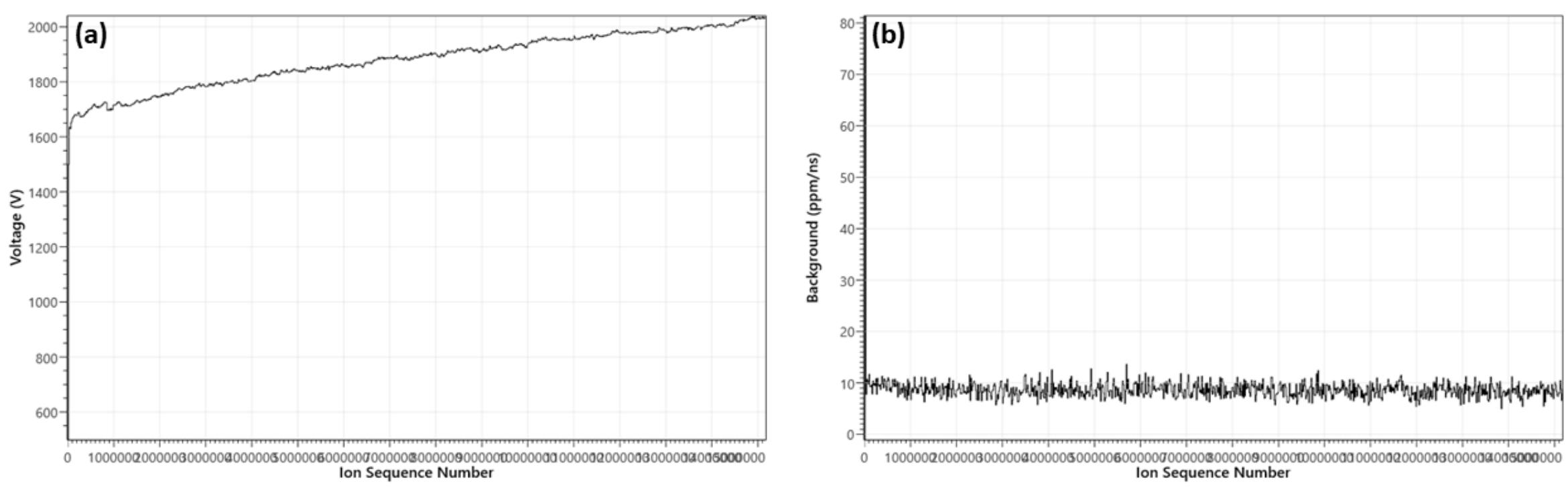

**Figure S11 –** (a) Representative voltage curve and (b) background from the atom probe tomography measurements. Both curves are stable, indicating the absence of artefacts induced by the experimental conditions during APT measurements.

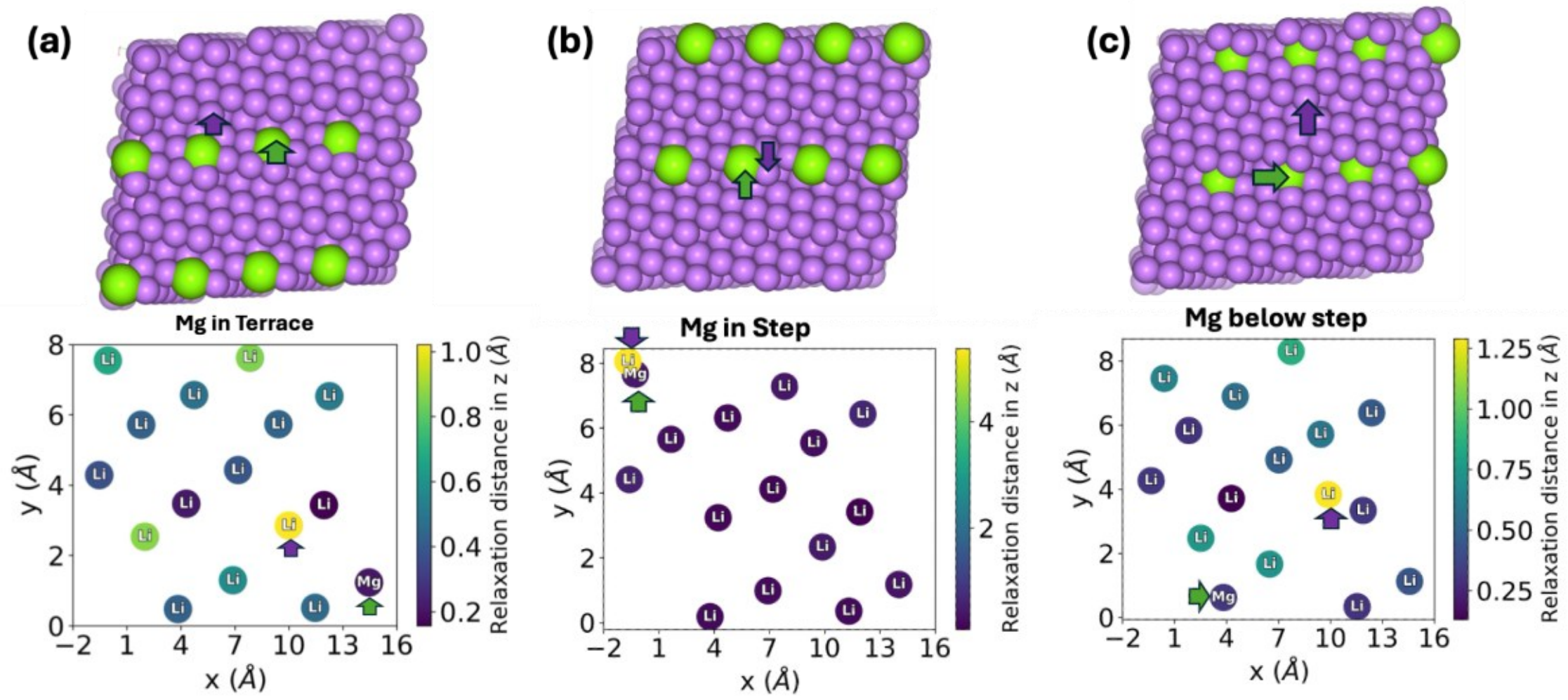

**Figure S12 –** Top row showing the starting unrelaxed configurations of three structure with (a) Mg in terrace, (b) Mg in step and (c) Mg below the step. The bottom row shows the x, y positions of top layer atoms, colored according to their relaxation distance in z directions. The Mg atom and the Li atom of interest are indicated with a green and purple arrow respectively.